# Thermodynamic, Kinetic and Mechanical Modeling to Evaluate $CO_2$-induced Corrosion via Oxidation and Carburization in Fe, Ni alloys

Aditya Sundar[a*], Aaron Feinauer[b], Bryan Kinzer[c], Joerg Petrasch[b], Liang Qi[a], Rohini Bala Chandran[c*]

[a]Department of Materials Science and Engineering, University of Michigan, USA, 48105
[b]Department of Mechanical Engineering, Michigan State University, USA
[c]Department of Mechanical Engineering, University of Michigan, USA, 48105
*Corresponding authors: *adisun@umich.edu, rbchan@umich.edu*

*Abstract*

A computational framework integrating thermodynamics, kinetics, and mechanical stress calculations is developed to study supercritical $CO_2$-induced corrosion in model Fe-based MA956 and Ni-based H214 alloys. Empirical models parametrized using experimental data show surface oxidation and sub-surface carburization for a wide range of thermodynamic conditions (800–1200 °C, 1–250 bar). CALPHAD simulations based on empirical models demonstrate higher carburization resistance in H214 compared to MA956 below 900 °C and through-thickness carburization in both alloys at higher temperatures. Finite element modeling reveals enhanced volumetric misfit-induced stresses at oxide and carbide interfaces, and its critical dependence on the carbide chemistry and concentration.

# 1. Introduction

Multicomponent Fe and Ni alloys are candidate structural materials in several existing and emerging technologies. One such thrust area is the development of high-temperature and corrosion-resistant materials for the deployment of supercritical carbon dioxide (sCO$_2$) based Brayton cycles in thermal, concentrated solar and nuclear power plants [1–5]. sCO$_2$ Brayton cycles are promising alternatives to conventional steam-based Rankine cycles. It facilitates larger energy conversion (thermal energy to electricity) efficiencies by enabling power cycle operation at higher temperatures (T $\geq$ 700 °C) compared to Rankine cycles (T $\leq$ 550 °C). Additionally, due to the lower critical point of CO$_2$ (7.4 MPa, 31 °C) compared to water (22.1 MPa, 647 °C) the entire power cycle can be designed to operate in a single (supercritical) phase. In this context, our study focuses on assessing the corrosion behavior of Fe- and Ni- based multicomponent alloys when exposed to CO$_2$ operating conditions with pressures up to 250 bar and temperatures up to 1200 °C.

Materials exposed to sCO$_2$ are subject to an environment that is both oxidizing and carburizing (Figure 1). Oxidation potential arises due to the thermal equilibrium between CO$_2$, CO, and O$_2$, and the driving force for carburization stems from the Boudouard reaction that produces graphitic carbon, C, and CO$_2$ from CO [6]. Several experimental studies have documented the formation of oxide phases on the surface of Fe/Ni alloys. Cr or Al rich alloys form corundum phase Cr$_2$O$_3$ or Al$_2$O$_3$ depending on the Cr/Al ratio [7–9]. An inner spinel (Ni(Al,Cr)$_2$O$_4$) type oxide is also known to form in Haynes 214 [9]. Carbide phase formation has also been experimentally reported in sCO$_2$-rich conditions in metallic alloys [4,10–15]. Different carbide phases can form depending on the local carbon concentration, including M$_{23}$C$_6$, M$_7$C$_3$ and M$_3$C$_2$, where M is the metal. These carbide phases have been shown to precipitate subsurface at the interface between the metal oxide and the bulk metal, and typically at metal grain boundary sites. At these sites, the precipitated carbides can act as stress concentrators and therefore accelerate microstructural crack propagation. Additionally, they can also lead to brittle failure because of limited plastic deformation in the surface oxides [11,16–20]. Therefore, there is a need to understand corrosion mechanisms due to both oxidation and carburization for alloys designed to operate with sCO$_2$ working fluids.

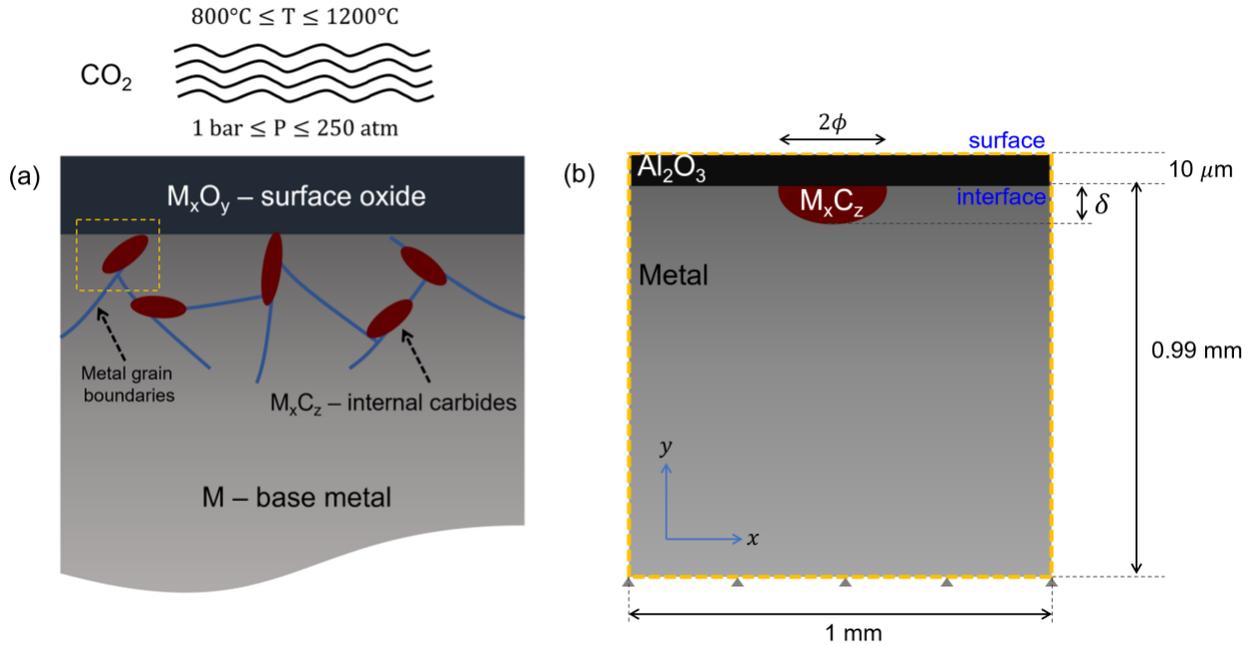

*Figure 1: (a) Schematic showing metallic alloy exposed to a $CO_2$ fluid environment; for temperature > 31 °C and pressure > 7.4 MPa, the state of the fluid is supercritical $CO_2$ ($sCO_2$). M denotes the base metal, $M_xO_y$ denotes the surface oxide and $M_xC_z$ denotes the carbide phases that form beneath the surface oxide. (b) Representative 2D finite element model used in this paper to model the impact of carbide phases on stresses generated in the oxide. $\phi$ and $\delta$ denote the lateral dimension and depth of the elliptical carbide precipitate.*

In recent studies, thermodynamic analyses have been performed using the CALPHAD (CALculation of PHAse Diagrams) method to probe the feasibility of carbide phase formation and to help interpret experimental observations [20–23]. Gheno et al. used thermodynamic equilibrium calculations to estimate the carbon activity at the interface between the metal and the metal-oxide, referred to as metal-oxide | metal interface hereafter, in steels [21]. Phase diagrams were obtained in this study by independently varying carbon activity and temperature, which demonstrated that the precipitation of metal carbide phases is thermodynamically feasible in Fe-9Cr. However, any effects of finite rates of species transport ($CO_2$, $CO$, $O_2$), and the influences of the oxide growth kinetics were precluded, and this limits the broad applicability of this study. To compute the carbon activity, $a_C$, thermodynamic equilibrium conditions for the Boudouard reaction were assumed, where the partial pressure of CO was dictated by the $O_2$ partial pressure obtained from the temperature-dependent $Fe/FeO_x$ oxidation reaction equilibrium from Ellingham diagrams [24]. It was also assumed that the fluid-phase partial pressure of $CO_2$ at the interface between the bulk metal and the metal oxide is equivalent to the $CO_2$ partial pressure at the surface. Discounting oxide growth kinetics and diffusion of gaseous species through the oxide can result in large variations in carbon activities as a function of the chemical composition of the oxide formed at the metal surface [4]. For instance, recent calculations estimate carbon activity in chromia ($Cr_2O_3$) forming alloys to be in the range of $10^9$–$10^5$ for operating temperatures of 550–800 °C respectively. At equivalent temperatures, this estimate becomes orders-of-magnitudes smaller, $10^{-8}$–$10^{-10}$ when iron oxide ($Fe_2O_3$) forming alloys were considered. This result is suggestive of $Cr_2O_3$ forming alloys being subject to a much larger driving force for carburization, dictated by larger carbon activity, as compared to $Fe_2O_3$ forming alloys at the same operating temperatures. However, this prediction is contrary to experimental observations wherein higher carburization is reported in $Fe_2O_3$-forming alloys. Several reported data points towards brittle carbide phases precipitating in relatively larger amounts in ferrous alloys (more likely to form Fe-containing alloys such as some $Fe_2O_3$) compared to nickel alloys (more likely to form $Cr_2O_3$), for similar temperature and pressure conditions [10,11,17]. Local carburization up to 12 wt% C was reported in AL-6XN stainless steel after a 3000 h exposure to $sCO_2$ at 650 °C, which promoted oxide spallation [10]. However, at similar conditions, no significant spallation was observed in nickel-based Haynes 230 alloys [10]. Similarly, $sCO_2$ exposure of SS316 at 650 °C and 20 MPa resulted in internal precipitation of chromium-rich carbides up to 100s of microns, with much less carburization reported in Ni-based Hastelloy-C276 at the same conditions [11]. Moreover, these calculations indicate a reduction in carbon activity with increasing temperature, which conflicts experimental observations of increased carburization at larger temperatures [4]. Hence, it is important to formulate a robust model that reflects experimental observations, while accounting for mechanisms of species transport.

From numerous experimental measurements including solubility, radiocarbon concentration, autoradiography and atom probe tomography, it is evident that the diffusion of elemental C through bulk oxides (FeO, $Fe_3O_4$, MnO, MgO, $Cr_2O_3$, $Al_2O_3$) is negligible [14]. In agreement with theoretical models [6,25], it is established that carbon transport from the surface to the metal-oxide | metal interface in $CO_2$ environments occurs via the diffusion of CO and $CO_2$ through grain boundaries or other similar nanoporous pathways in the surface oxide [6,14,25–28]. To include these mechanisms, few studies have proposed integrated thermodynamic and kinetic models to compute the interfacial carbon activity [6,14]. Rouillard et. al developed physics-based formalisms to provide insights that can explain experimental observations by predicting higher carburization potential at higher temperatures and pressures [6]. While their proposed model accounted for physically relevant mechanisms of oxide growth kinetics and gas species transport through the oxide layer, the carburization potential was not numerically calculated as a function of state conditions (temperature, pressure) and kinetic parameters (oxidation rate, diffusion rate). Therefore, it is necessary to develop modeling tools that can account for the coupled effects of the kinetics of metal-oxide growth and species transport ($CO_2$, CO, and $O_2$) through the oxide layer, to yield realistic predictions for the carburization behavior in $sCO_2$ conditions.

As experiments have demonstrated, carbide precipitation is more severe in Fe alloys compared to Ni alloys, under identical state conditions [10,17]. While it is hypothesized that severe carburization near the metal-oxide | metal interface promotes oxide spallation or delamination in these materials, a systematic investigation of the impact of carbide precipitates on the mechanical stability of surface oxides has not been performed. Residual stresses can arise due to volumetric misfits between the matrix and precipitate phases [29–33] which results in stress accumulation at the metal-oxide | metal interface. While metals can absorb these stresses via plastic deformation, increasing residual stresses due to large carbide inclusions can promote the initiation and propagation of cracks within the oxide, leading to eventual oxide delamination from the metal surface [34–37]. Therefore, there is also a need to connect the extent and chemistry of carbide precipitation to the evolution of stress fields at the various interfaces.

Motivated by these knowledge gaps, the main objective of this study is to develop an integrated thermodynamic and kinetic model that is parametrized using experimental data to evaluate oxidation and carburization behavior of Fe- and Ni-based alloys in $CO_2$ environments (1–250 bar $CO_2$; 550–1200 °C). A reaction-diffusion model is formulated to compute the driving force for carburization by considering temperature and oxygen partial pressure dependent kinetic parameters for oxide growth rates, and finite rates of species diffusion through the metal oxide layer. Carbon activity computed from this model is coupled to CALPHAD calculations to present comparative analyses of carburization behavior for the Fe and Ni alloys. A secondary objective of this study is to assess the effects of carbide precipitates on mechanical stability of the various interfaces between the metal-oxide, metal, and the metal-carbide. To this end, we've developed a new approach that couples calculations of residual stress due to volumetric misfits with CALPHAD calculations for the extent and composition of carbide precipitation for the two alloys. Distinct from what has been reported in prior work, our study: (1) advances modeling tools to provide quantitative predictions for interfacial carbon activity over a wide range of temperatures and pressures, which is relevant to materials and components for next generation thermal technologies operating in a $CO_2$ environment; (2) evaluates carburization depths and carbide precipitate compositions in model Fe and Ni alloys; and (3) reveals the impact of carbide chemistry and precipitate morphology (quantified by its size and aspect ratio) on residual stresses at the interface between the metal-oxide and metal-carbide.

## 2. Methods

### 2.1 Materials

Two different materials are evaluated – (1) Fe-based Incoloy MA 956 and (2) Ni-based Haynes 214 (H214) when exposed to $CO_2$ conditions. Their chemical compositions are listed in Section S1 of the Supplementary Information (SI) [38,39]. Incoloy MA 956 is an attractive candidate for high-temperature applications in oxidizing environments, due to its superior strength and creep properties. MA 956 is reported to have a Youngs modulus of ~70 GPa and a 1000-hour creep strength of ~70 MPa at temperatures as high as 1000 °C [38]. In a carburizing environment, Fe-based alloys face the shortcoming of enhanced carbide precipitation [10,17]. However, Ni-based H214 alloys demonstrate improved high temperature corrosion resistance at the expense of mechanical strength [39] (1000-hour creep strength of 2-8 MPa at 1000 °C) [39].

### 2.2 Thermodynamic Phase Equilibrium Calculations

The CALPHAD method is applied to determine the thermodynamically stable system configuration, including chemical composition and mass/mole fractions of equilibrium phases, as a function of state conditions [40]. The total free energy of the system is expressed as the weighted sum over of the molar Gibbs free energy over all phases, as a function of temperature, pressure, and phase composition. The molar Gibbs free energy for each phase includes the following contributions: 1) reference energy obtained from the elements that constitute each phase; for instance, the reference states for Fe and Ni are body centered cubic (BCC) and face centered cubic (FCC) crystal structures respectively, 2) configurational entropy, 3) physical phenomena such as magnetism and 4) excess free energy [40]. Detailed formulations for the temperature and pressure dependencies of the energy contributions are described in the ref. [40]. The Thermo-Calc software [41] was used for all CALPHAD-based predictions for equilibrium phases reported in this manuscript. Particularly, TCFE11 Steels/Fe-alloys, MOBFE5 Steels/Fe-alloys, TCNI10 Ni-alloys and MOBNI5 Ni-alloys databases were used for equilibrium predictions. These databases indicate that the compositions modeled in this study (Table 1) are within recommended limits for individual alloying element [42].

## 2.3 Model for Oxidation and Carburization in sCO₂ Environments

When a material is exposed to sCO₂ fluids there are two reactive/corrosive mechanisms at play — (1) oxidation, which results in an oxide layer that forms and passivates the surface, and (2) carburization that occurs internally within the bulk alloy. The pertinent reaction equilibria to consider includes the CO₂ dissociation reaction (Equation 1) and the Boudouard reaction (Equation 2). Because CO₂ can exist in a gaseous or a supercritical fluid state depending on state conditions, for the sake of generality, Equations 1, 2 denote CO₂ with a fluid (*f*) state.

$$CO_2(f) \leftrightharpoons \frac{1}{2} O_2(g) + CO(g) \tag{1}$$

$$2CO(g) \leftrightharpoons C(s) + CO_2(f) \tag{2}$$

The driving force for surface oxidation stems from CO₂ dissociation equilibrium (Equation 1). Oxygen generated in this step oxidizes the metal M, up to a stoichiometric coefficient (not included in Equations 3):

$$2Al(s) + \frac{3}{2} O_2(g) \leftrightharpoons Al_2O_3(s) \tag{3}$$

The net metal oxidation reaction in a CO₂-environment can be expressed as Equation 4 which is a combination of Equations (1) and (3):

$$2Al(s) + 3CO_2(g) \leftrightharpoons Al_2O_3(s) + 3CO \tag{4}$$

Based on experimental data, Al₂O₃ can be assumed to be the only surface oxide formed for temperatures ranging from 800–1200 °C [8,43]. To model different oxides, the stoichiometric coefficient can be changed accordingly in Equation 4. As will be shown in Section 3.1, the carbon activity resulting from the thermal dissociation and the Boudouard reactions (Equations 1 and 2) is insufficient to result in carbide phase formation at the surface ($a_C \sim 10^{-10}$), where the oxygen activity dominates. However, significantly large carbon activities (up to $10^{-3}$) at the metal-oxide | metal interface can result in carbide precipitation. Carbon activity, $a_C$, at this interface is calculated by quantifying the coupled effects of diffusion of carbon containing species, CO₂ and CO, through the oxide and the kinetics of oxide growth. Key assumptions in the formulated model are: (1) quasi-steady, diffusion controlled parabolic rate of metal oxidation, (2) diffusion of elemental C through the metal oxide layer is assumed to be negligibly small and justified by the diffusivity data [14,26,27], and (3) the progress of metal oxidation via Equation 4, where $CO_2$ is the source of $O_2$ leading to metal-oxide and CO formation.

At quasi steady-state, there is no accumulation of any species at the metal-oxide | metal interface, and the net diffusive species flux is balanced by the net reaction rate (per unit surface area), i.e., consumption of $CO_2$ and production of CO, resulting in Equation 5,

$$J_i = r_i; \; i = CO_2, CO \tag{5}$$

where, $J_i$ is the molar diffusive flux and $r_i$ is the molar reaction rate normalized per unit surface area, i.e., units of mol/m²/s. $J_i$ is obtained from Fick's Law as in Equation 6,

$$J_i = \frac{D_{i,\text{eff}}\left(c_i^{\text{surf}} - c_i^{\text{int}}\right)}{x_{\text{oxide}}}; \; i = CO_2, CO \tag{6}$$

where, $D_{i,\text{eff}}$ is the effective diffusivity of the fluid species $i$ through the oxide layer; $c_i^{\text{surf}}$ and $c_i^{\text{int}}$ denote $i^{\text{th}}$ species concentrations at the surface and the metal-oxide | metal interface respectively; $x_{\text{oxide}}$ is the metal-oxide layer thickness. At the surface, it is assumed that $P_{CO}^{\text{surf}} \cong 0$, which is justified based on the $CO_2$ mole fraction being close to 1 for the temperatures considered (from Equation 2).

The effective diffusivity in Equation 7a is modeled as a function of the bulk binary CO-$CO_2$ diffusivity and microstructural features of the oxide layer, including, porosity, $\epsilon$, and tortuosity, $\xi$. The binary diffusivity is modeled using the method of Fuller et al., which is derived from the Chapman-Enskog theory [44], However, the Chapman-Enskog model is not directly applicable to high pressure conditions (non-dilute gas regime). To account for reduced diffusivities at high pressures, an empirical correction is introduced as shown in Equation 7b. Since $D \propto \lambda^2$ and $\lambda \propto P^{-1}$, (where $\lambda$ is the mean free path), an empirical correction of the form $\left(\frac{P_{\text{ref}}=1\text{ bar}}{P}\right)^2$ is introduced and shown in Equation 7b, which results in reduced diffusivities at high pressures. In Equations 7b-c, $M_i$ and $V_i$ are the molecular mass and equivalent volume of species $i$. Knudsen diffusivity is not included because for most of the pressure and temperature conditions modeled in this work, the mean free path is anticipated to be much smaller than representative pore-size of 25 nm (shown in Section S2 of the SI). 25 nm is used as baseline value since reported grain boundary channel widths in chromia grown on Fe-20Cr after $CO_2$ exposure are tens of nanometers from tomography data [45].

$$D_{i,\text{eff}} = \frac{\epsilon}{\xi} D_{ij} \tag{7a}$$

$$D_{ij} = \frac{0.00143 T^{1.75}}{p M_{ij}^{0.5} \left[V_i^{\frac{1}{3}} + V_j^{\frac{1}{3}}\right]^2} \left(\frac{P_{\text{ref}} = 1 \text{ bar}}{P}\right)^2 \tag{7b}$$

$$M_{ij} = \frac{2}{\frac{1}{M_i} + \frac{1}{M_j}} \tag{7c}$$

Porosity and tortuosity will be influenced by the size, concentration, and distribution of grain boundaries in the oxide layer, with additional influences from temperature. While specific values for $\epsilon$ and $\xi$ are challenging to glean from reported experimental data, we parametrize the ratio of porosity to tortuosity, i.e., $\gamma = \frac{\xi}{\epsilon}$, with a baseline value of 47.5 for this ratio. This value was tuned such that the $a_C = 1$ at 1000 °C and $P_{CO_2}^{\text{surf}} = 250$ bar. While these microstructural parameters could have intrinsic temperature dependencies, our model assumes a fixed $\gamma$ value for all thermodynamic conditions.

The area-specific molar rate of $CO_2$ consumption and CO production, $r_i$ (introduced in Equation 5) is governed by a quasi-steady rate of metal-oxide formation, and is defined in Equation 8. This formulation considers the stoichiometric balance between moles of $CO_2$ consumed and moles of CO generated, as formulated in Equation (4). Details are provided in Section S3 of the SI.

$$r_i = \frac{3(k_{\text{oxide}}^* t)^{\frac{1}{2}}}{2 M_{\text{oxide}} t}; i = CO_2, CO \tag{8}$$

$$k^*_{\text{oxide}} = k_{\text{oxide}} \times \left( \frac{c^{\text{int}}_{\text{CO}_2}}{c^{\text{int}}_{\text{CO}_2} + c^{\text{int}}_{\text{CO}}} \right)^{0.55} \tag{9}$$

$$k_{\text{oxide}} = k_0 \exp\left(-\frac{E_a}{RT}\right) \left(\frac{P^{\text{surf}}_{\text{ox}}}{P_{\text{ref}}}\right)^c, \text{ox} = \text{CO}_2, \text{air}, \text{water vapor}; 800\,°C \leq T \leq 1200\,°C \tag{10a}$$

$$k_0 = 0.924\ (0.027, 31.336)\ \text{mg}^2 \text{cm}^{-4} \text{s}^{-1}; \tag{10b}$$

$$E_a = 1.512 \times 10^4 (1.152 \times 10^4, 1.873 \times 10^4)\ \text{J mol}^{-1};$$

$$c = 0.55$$

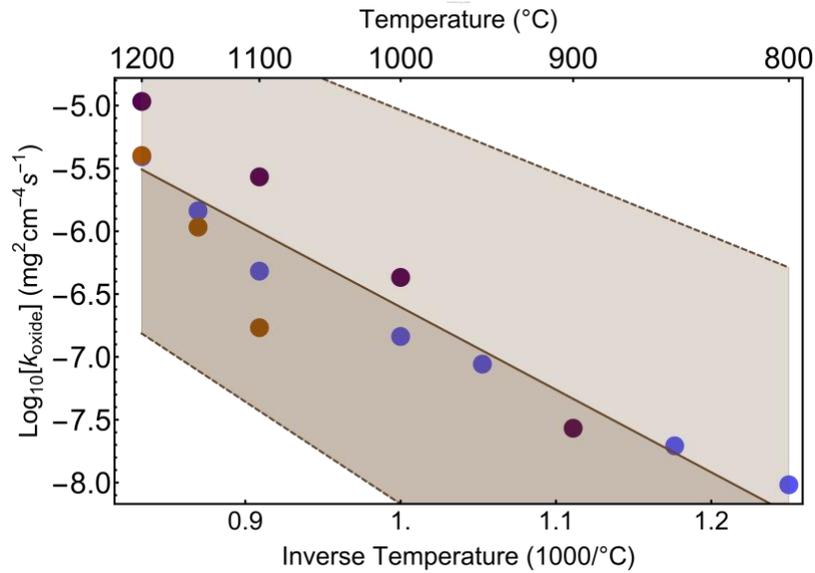

*Figure 2: Temperature dependence of the parabolic oxidation rate. Different colors for the data points denote different alloys. The central solid brown line is the ordinary least squares linear fit to the data. The shaded regions between the upper and lower dashed brown lines denote the 95% confidence interval.*

In Equation 8, $k^*_{\text{oxide}}$ is a concentration modified parabolic rate constant, $t$ is the time elapsed since the onset of oxide growth and $M_{\text{oxide}}$ is the molar mass of the oxide formed. In our calculations, molar masses of alumina (102 g/mol) and chromia (152 g/mol) are considered depending on the alloys modeled. For MA956 and H214, $Al_2O_3$ is reported to be the dominant surface oxide for 800 °C < T < 1000 °C [7,8]. Whereas, for SS316H and Ni740 (alloys that have larger amounts of Cr) $Cr_2O_3$ is the surface oxide for 400 °C < T < 800 °C [46]. Equation 9 obtains $k^*_{\text{oxide}}$ as a function of the fitted rate constant, $k_{\text{oxide}}$, deduced from experimental corrosion measurements and the mole fraction of the $CO_2$ species present at the metal | metal-oxide interface. This mole fraction is raised to the power of 0.55, which was published in Holcomb et al., by fitting parabolic oxidation rates with pressure at 700 °C [13].

The fitted rate constant, $k_{\text{oxide}}$, is only a function of experimentally accessible temperature and pressure conditions at the surface of the metal, and obtained using Equation 10a, where T is the temperature in °C and $P_{CO2}^{\text{surf}}$ is the total pressure of $CO_2$ at the surface in bar and the reference pressure, $P_{\text{ref}}$ is 1 bar. While the absolute $k_{\text{oxide}}$ values strongly depend on the material composition, the trend in its temperature and pressure dependence is applicable more generally for many compositions [12,13,47]. Published $k_{\text{oxide}}$ data for MA956, FeCrAl and H214 alloys in an air environment over 800–1200 °C was used [8,9]. An Arrhenius rate equation form (Eq. 10(a)) can be fitted to these datasets, with a constant pre-factor, $k_0$, and an exponential activation term that decreases with increase in temperature with a barrier potential of, $E_a$. Best-fit values for $k_0$ and $E_a$ are shown in Equation 10(b) along with their 95% confidence intervals, and point to large variations in the pre-factor term driven by the large scatter in the experimental $k_{\text{oxide}}$ data. Figure 2 shows the reported oxidation rate constant data as a function of temperature and corresponding Arrhenius fit ($R^2 = 0.9$). In addition to the temperature dependence, the pressure dependence of $k_{\text{oxide}}$ was obtained in Equation 10a from experimental data reported by Holcomb et al., for Ni-based alloys exposed to $CO_2$ at 700 °C over a pressure range of 1–240 bar [13]. While the fitting quality is suboptimal ($R^2 = 0.33$), this relationship is used as such due to the lack of other high-quality measured data. As will be shown in the results (Section 3.2), the deduced fit for $k_{\text{oxide}}$ in Equation 10a compares reasonably well with also in-house experimental measurements for H214 coupons exposed to $CO_2$ in a corrosion rig at $P_{CO_2} = 25$ bar and 1100 °C.

By modeling the Boudouard reaction in Equation 2 to be at equilibrium, the carbon activity, at the at the metal-oxide | metal interface, $a_C^{\text{int}}$, can be calculated using Equation 11,

$$a_C^{\text{int}} = K_B \frac{\left(\frac{P_{CO}^{\text{int}}}{P_{\text{ref}}}\right)^2}{\frac{P_{CO2}^{\text{int}}}{P_{\text{ref}}}} \tag{11}$$

where, $K_B$ is the temperature dependent equilibrium constant for the Boudouard reaction (Equation 12), and applicable for temperatures in the range of 227–1927 °C [48]. $K_B$ decreases with increase in temperature due to the reduction in the driving force for solid carbon formation at higher temperatures.

$$\text{Log}_{10}[K_B] = \frac{9141}{T} + 0.000224 \times T - 9.595 \quad (12)$$

In Equation (13), $P_{CO}^{int}$ and $P_{CO2}^{int}$ are obtained from their respective concentrations, $c_{CO}^{int}$ and $c_{CO2}^{int}$, and by applying by the van der Waals equation of state provided in Section S4 in the SI [49]. The carbon activity computed from Equation 11 is used as a boundary condition to perform 1-D diffusion simulations paired with equilibrium calculations for carbide phase precipitation in the DICTRA module of Thermo-Calc [50]. A brief description of the DICTRA method is provided in Section S5 of the SI. The matrix phase for Incoloy MA 956 and Haynes 214 is modeled as body centered cubic (BCC) and face centered cubic (FCC) respectively. The thermodynamic description of the following carbide phases is included: $M_{23}C_6$, $M_7C_3$, $M_5C_3$ and $M_3C_2$. Implementation details for these simulations are discussed in literature [41].

## 2.4 Finite Element Analysis to Understand Oxide Fracture

*Formulation*: Strain misfits between different phases, including the base metal, metal-oxide, and metal-carbide precipitates, can induce stresses in the ceramic phases due to their higher elastic moduli compared to metals. Equation 13 provides the strain misfit obtained due to specific volume mismatch between the precipitate and the matrix phase,

$$\epsilon_{v,im} = \frac{(v_0^i)^{\frac{1}{3}} - (v_0^m)^{\frac{1}{3}}}{(v_0^m)^{\frac{1}{3}}}; \quad i = Cr_{23}C_6, Cr_7C_3, Cr_3C_2; \quad m = \text{MA956, H214} \quad (13)$$

where, $\epsilon_{v,im}$ is the strain misfit, $v_0^m$ is the volume per metal atom of the matrix and $v_0^i$ is the volume per metal atom of the precipitate $i$ [32]. To simulate the stresses induced by these misfit strains, a heat-treatment process is modeled with a fixed thermal loading of $\Delta T = 1$ °C. Each phase is assigned an equivalent thermal expansion coefficient, $\alpha_v$, which is related to the strain misfit as shown in Equation 16, from reference [33]. With this formulation, a thermal loading step with $\Delta T$ change in temperature will introduce residual stresses due to the misfit strains.

$$\alpha_{v,im} = \frac{\epsilon_{v,im}}{\Delta T}; \quad (14)$$

*Material Properties*: The misfit strains and isotropic expansion coefficients assigned to the carbide, metal and the metal oxide phases are listed in Section SX of the SI. Three carbide precipitates were modeled — $Cr_{23}C_6$, $Cr_7C_3$ and $Cr_3C_2$ —based on predicted compositions from phase equilibrium calculations in CALPHAD (Section 3.2). $v_0^i$ and $v_0^m$ values are calculated using data from literature that were computed via density functional theory [51–53]. $v_0^{MA956}$ is approximated as $v_0^{APMT}$ since reference data is only available for the commercially available Fe-alloy, APMT (Fe-21Cr-5Al); this alloy has Cr and Al concentrations like that of MA956, and therefore this assumption is reasonable [52]. $v_0^{H214}$ is obtained by interpolating data for pure Ni and $Ni_{0.75}Cr_{0.25}$ [53] . Even though the misfit strains are expected to vary with temperature due to thermal expansion, these values are assumed to be independent of temperature for the sake of simplicity. The strain misfit for the metal alloys relative to itself is assigned a reference isotropic value of 0 (Table 2). In both sets of simulations (for MA956 and H214 alloys), it was assumed that the oxide is only strained parallel (∥) to the metal-oxide | metal interface ($y$ axis in Figure 1b). The strains in the oxide perpendicular (⊥) to the metal-oxide | metal interface ($x$ axis in Figure 1b) are set equal to 0, assuming that the oxide is not constrained in the surface normal direction [54]. Strain misfits for the MA956 | $Al_2O_3$ and H214 | $Al_2O_3$ interfaces were obtained from prior literature [34,54]. For the internally precipitating carbide phases, the corresponding volumetric misfits were calculated using the molar volumes (per metal atom) for the matrix and precipitate phases (denoted $v_0^m$ and $v_0^m$ in Equation 13) [32]. The carbides are assumed to expand isotropically.

*Implementation:* A 2-D modeling domain (Figure 1b) was used to perform finite element analysis in Abaqus [55] to compute residual stresses due to volumetric misfit strain. This strain is obtained using the pseudo thermal expansion coefficient (Equation 14). To apply a thermal load and generate the stress profile, the entire domain was subjected to a $\Delta T = 1$ °C consistent with how the strain misfit was defined in Equation 14 (this ensures a step of 1 °C; the exact temperature doesn't affect the results). The metal was modeled as a 0.99 mm × 1 mm domain with a 10 μm thick oxide layer. Using our formulation for $k_{oxide}$ in Equations 10a & 10b, the calculated film thickness at 1100 °C, 250 bar $CO_2$ after 100 h is ~6.7 $\mu$m. The calculated film thickness at 1200 °C, 250 bar $CO_2$ after 100 h is ~14.5 $\mu$m. Therefore, a 10 $\mu$m oxide layer is modeled, which is representative of steady oxide growth for ~100 h of material exposure to $sCO_2$ environments.

A single semi-elliptical precipitate was modeled for the carbide phase at the metal-oxide | metal interface with varying sizes parametrized by the lateral dimension $\phi$ and the normal dimension $\delta$; $\phi$ was varied between 1 and 100 $\mu$m, and the aspect ratio, $\phi/\delta$, was varied from 1 to 5. We model

an individual precipitate at the metal-oxide | metal interface as a two-dimensional representative volume element, to probe effects of carbide precipitation on oxide layer delamination. In this model, the precipitate is aligned with its major axis along the metal | oxide interface, $i.e.,$ y-axis in Figure 1b. This is a simplification compared to reality where the carbide precipitates will be present throughout the bulk of the alloy, especially along grain boundaries which are sites for heterogeneous nucleation. While internal carbide precipitates can enhance the mechanical strength of the alloy, carbides near the oxide layer can introduce large tensile stresses in the oxide and lead to brittle failure and subsequent oxide spallation. Therefore, our modeling domain is designed to study the impact of carbides near the metal-oxide | metal interface which can promote oxide failure. Additionally, a semi-elliptical precipitate shape is considered as it maximizes the contact area with the surface oxide. Stresses generated in the oxide for this configuration will be higher compared to stresses generated for more spherical precipitate orientations. Therefore, these simulations are representative of a worst-case scenario for oxide fracture.

For the boundary conditions, the bottom edge of the domain was constrained along the normal/y-axis. The top and lateral edges were unconstrained to allow deformation due to the misfit strains imposed by the carbide precipitate. Meshes are generated in this domain with a dominant number of quadrilateral elements (> 99.9% in a total of more than 50,000 elements) with an average size of 4 $\mu$m × 4 $\mu$m along $x$ and $y$ dimensions for the base metal; within the oxide layer the mesh is more refined and each element roughly has a size of 4 $\mu$m × 0.5 $\mu$m; within the carbide precipitate nodes are located at a spacing of 4 $\mu$m along the periphery. Mesh convergence studies were performed to indicate that the maximum principal stress computed from the model changes by only ~1% with a more refined mesh size of 0.4 $\mu$m × 0.4 $\mu$m within the base metal and oxide domains.

The stress components are computed via the constitutive Equations 15a-e,

$$\sigma_x = \frac{(\epsilon_x + \nu\epsilon_y)E}{(1 - \nu^2)} \tag{15a}$$

$$\sigma_y = \frac{(\epsilon_y + \nu\epsilon_x)E}{(1 - \nu^2)} \tag{15b}$$

$$\tau_{xy} = \frac{E\gamma}{2(1 + \nu)} \tag{15c}$$

$$\epsilon_x = \frac{\Delta x}{x}, \epsilon_y = \frac{\Delta y}{y}, \gamma = \epsilon_{xy} + \epsilon_{yx}, \epsilon_{xy} = \frac{\Delta x}{y}, \epsilon_{yx} = \frac{\Delta y}{x} \qquad (15d)$$

$$\sigma_1 = \frac{\sigma_x + \sigma_y}{2} + \sqrt{\left(\frac{\sigma_x - \sigma_y}{2}\right)^2 + \tau_{xy}^2} \qquad (15e)$$

where, $\sigma_x$ and $\sigma_y$ denote the normal stresses along $x$ and $y$, $E$ is the temperature-dependent Young's modulus, $v$ is the Poisson ratio, $\tau_{xy}$ is the shear stress, $\gamma$ denotes the total shear strain, $\epsilon_x$ and $\epsilon_y$ are the normal strains computed using the thermal expansion coefficients as described in Equation 14, and $\sigma_1$ is the maximum principal stress which is used to evaluate the likelihood of oxide fracture by comparing with experimental fracture stresses. For plastic deformation in the metal domain, the input data for plastic stress-plastic strain is used to obtain the stress values at different strains. The stress-strain relations in Equations 15a-e are valid in the elastic deformation regime. The metallic phases MA956 and H214 are assumed to deform elastically and plastically, whereas $Al_2O_3$ and the carbide phases are assumed to deform only elastically. This is a reasonable assumption since ceramics generally have larger elastic moduli, higher yield strengths, and much less dislocations (and other plastic deformation defects) compared with metals due to strong ionic bonds. Temperature dependent plastic deformation parameters for MA956 were obtained by digitizing published data for FeCrAl alloys [56]; the same parameters were used for H214 due to the lack of deformation data. Plastic deformation data at the highest available temperature (1014 K) was used for conservative estimates of stresses. Temperature dependent Young's moduli for MA956, H214 and $Al_2O_3$ were obtained from refs. [57–59] and for the carbide phases from refs. [60–62] (Section S6 in SI). Poisson ratios $(v)$ for the base metals and the carbides are assumed to be independent of temperature and these values are obtained from refs. [60].

## 3. Results and Discussion

### 3.1 Surface Oxidation in a $CO_2$-Environment

Figure 3a shows the equilibrium activities of $O_2$ and C from Equations 1 and 2, corresponding to 1 and 250 bar $CO_2$ total pressure at the surface of the alloy. For the entire range of temperature and pressure modeled, the oxygen activity, $a_{O_2}$, is significantly larger by several orders of magnitude than the carbon activity, $a_C$, at the surface. This indicates a substantial driving force for oxidation reactions occur at the surface. While there is weak carbon activity at the surface, carbon activity will increase with depth into the bulk metal as the oxygen gets consumed to form the metal-oxide (Figure 1). Therefore, preferential carbide precipitation is expected to occur below the metal-oxide layer. The oxygen activity decreases, by roughly one order-of-magnitude, with increase in total $CO_2$ pressure dictated by the thermodynamic equilibrium of $CO_2$ dissociation (Equation 1) shifting to the left when pressure increases. A contrasting trend is observed for carbon activity, which increases as $CO_2$ pressure increases, from ~$10^{-11}$ to ~$10^{-10}$ at 1200 °C.

Figure 3b shows the phase diagram for the oxides formed for a general FeCrAlTi alloy, whose composition is the same as MA956 (Table 1) without other alloying elements, as a function of temperature and oxygen activity. Larger oxygen activities and higher temperatures favor the formation of corundum ($M_2O_3$) and rutile ($MO_2$) phases, where M is the metallic element, since they have large oxygen concentrations per metal atom. For example, the rutile phase is $TiO_2$ across the 800–1200 °C range, at $a_{O_2} = 10^{-2}$. Under the same conditions, the corundum phase is $Fe_{1.16}Cr_{0.58}Al_{0.26}O_3$. Thermodynamically, this indicates that Fe, Cr and Al can all be oxidized to form corundum even when exposed to a $CO_2$ environment at high temperatures. While the thermodynamically predicted corundum also contains Fe, the real composition will be influenced by other kinetic factors such as diffusivity of the oxidizing element in the base alloy, diffusivity of oxygen through the oxide, etc. The spinel phase, with a smaller oxygen concentration per metal atom, is stable when oxygen activity is poor. For instance, at a lower value of $a_{O_2} = 10^{-8}$, the spinel composition is $Fe_{1.18}Cr_{0.56}Al_{0.26}O_4$. While the general composition of equilibrium oxide phases can be predicted from these calculations, the specific chemical composition and stoichiometry of the oxide phases is determined by kinetic parameters such as ionic diffusion through the metal and oxide phases, volumetric mismatch between the oxides and metal (quantified as the Pilling-Bedworth ratio or P-B ratio, $r_{P-B}$ [63]). Defined as the ratio of the volume per metal atom of the oxide and the volume per atom of the corresponding metal, $1 < r_{P-B} < 2$ is likely to result in a passive oxide coating, observed in Al ($Al_2O_3$), Ti ($TiO_2$), Cr ($Cr_2O_3$), etc. In metals like Mg, $r_{P-B} < 1$ result in thin and partial oxide layers. In metals like Fe, $r_{P-B} > 2$ promotes the spallation of oxide coatings and eventual rusting. The $Fe_{1.16}Cr_{0.58}Al_{0.26}O_3$ oxide predicted to form at $a_{O_2} = 10^{-2}$ has a composition weighted $r_{P-B}$ value of 1.86 and is likely to be passivating in oxidizing atmospheres.

Figures 3c and 3d show the equilibrium carbide phases as a function of carbon activities and temperatures for MA956 and H214. Carbon activity, $a_C$, in Figures 3c, 3d have a minimum value of $10^{-5}$, which is many orders larger than that predicted at the metal surface in Fig. 3(a). This is done to capture the effects of carburization beneath the metal-oxide layer, where a_c will be relatively large compared to the surface. MA956 is primarily BCC with carbides of type $M_{23}C_6$ and $M_7C_3$ stable carbon activities larger than ~$10^{-3}$. At 800 °C and $a_C = 10^{-2}$, the carbide composition is $Cr_{18.63}Fe_{4.37}C_6$. At 800 °C and $a_C = 10^{-1}$, the carbide composition is $Cr_{5.67}Fe_{1.33}C_3$. Similarly, H214 is primarily FCC with $M_{23}C_6$, $M_7C_3$ and $M_3C_2$ carbides stable depending on the carbon activity and temperature values. For instance, at 800 °C, the carbide phases are $Cr_{6.54}X_{0.46}C_3$ (X = Mo, Fe, Ni) and $Cr_3C_2$ (no other alloying elements) at $a_C = 10^{-0.5}$. For lower carbon activities, $a_C = 10^{-2}$, $Cr_{20.22}X_{2.78}C_6$ (X = Mo, Fe, Ni) is stable with a lower carbon concentration. Generally, the stoichiometric carbon content in the carbide phases increases with increasing $a_C$ values. At higher temperatures, larger carbon activity values are required to stabilize the carbide phases. For instance, at 1200 °C, no carbide phases are stable in H214 even at $a_C = 10^{-2}$ (as seen in Figure 3d). Since the solubility of C in Ni increases with temperature, higher carbon activities are needed to precipitate carbide phases out of the Ni-matrix.

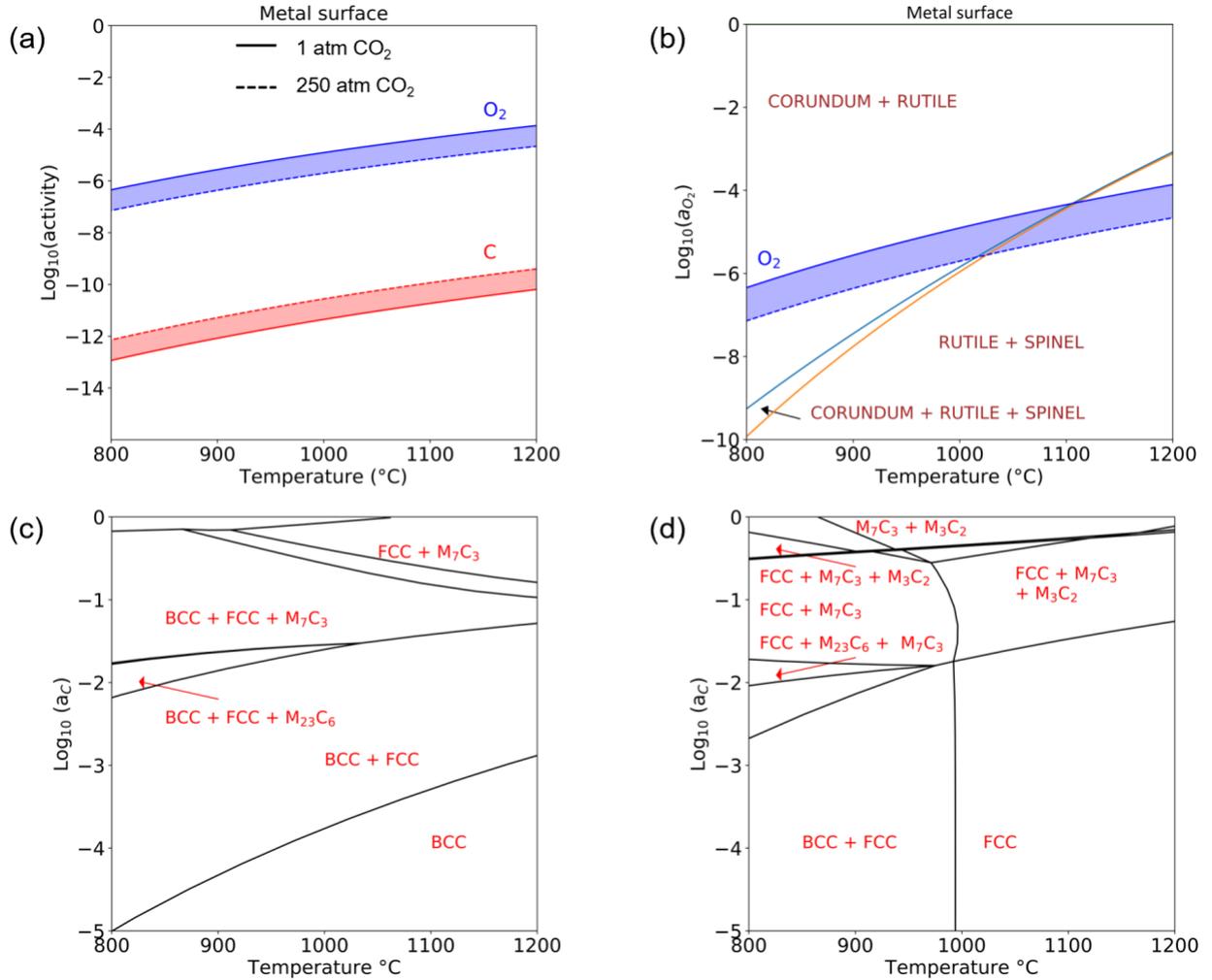

Figure 3: Thermodynamic phase diagram calculations using CALPHAD. (a) Equilibria between Equations 1 and 2. (b) shows the phase diagram for FeCrAl with temperature and a(O₂) as independent variables. (c) and (d) show phase diagrams for MA956 and H214 with temperature and $a_c$ as the independent variables.

## 3.2 Carburization in a $CO_2$-Environment

## Experimental determination of $k_{oxide}$

The deduced dependency of $k_{\text{oxide}}$ on P and T (Equation 10a) was compared against in-house experimental measurements performed in a corrosion rig designed at Michigan State University. Corrosion experiments were performed in research-grade $CO_2$ (Airgas, 99.999% purity) at 1100 °C and 25 bar pressure to track the transient change in mass of H214 metal coupons. More details about the corrosion chamber design and operation to maintain high pressure and temperatures are presented in Section S7 in the SI. Additively manufactured H214 coupons were electro-discharge machined to a size of 8 mm × 8 mm × 1.5 mm with a 2 mm hole near the edge to allow hanging of the sample during exposure. 18 samples were hung from the top of the isothermal zone in the corrosion chamber via solid alumina rods and separated with alumina spacers (Figure 4a). The specimens were wet ground with 1200 grit silicon carbide paper on all surfaces and 800 grit on all sides. Samples were cleaned in ethanol prior to the tests and measured using a micrometer with +/- 0.001 mm accuracy. The exposed samples were weighed on a scale with +/- 0.1 mg accuracy.

Figure 4b shows the measured mass change values averaged over 3 samples after several hours of exposure to $CO_2$ in the corrosion chamber. The measured data shows trends for transient mass changes that are consistent with a parabolic growth model for oxidation on the metal surface (Equation 10a). The decrease in experimental mass-change values, for instance, at 200 h and 350 h, can be attributed to surface oxide spallation. These results are in the same order-of-magnitude as the mass change from $k_{\text{oxide}}$ (Equation 10a) obtained from prior experimental data, which was obtained over an extended range of temperatures. Therefore, the deduced $k_{\text{oxide}}$ from prior data was used, in place of the measured value, for further calculations, that is restricted to a specific oxide and based on measurements only at 1100 °C. Overall, our approach to model parabolic oxide growth rates and data-driven $k_{\text{oxide}}$ dependence on T and P is reasonable and can be applied with confidence to determine interfacial carbon activity and carbide precipitation effects.

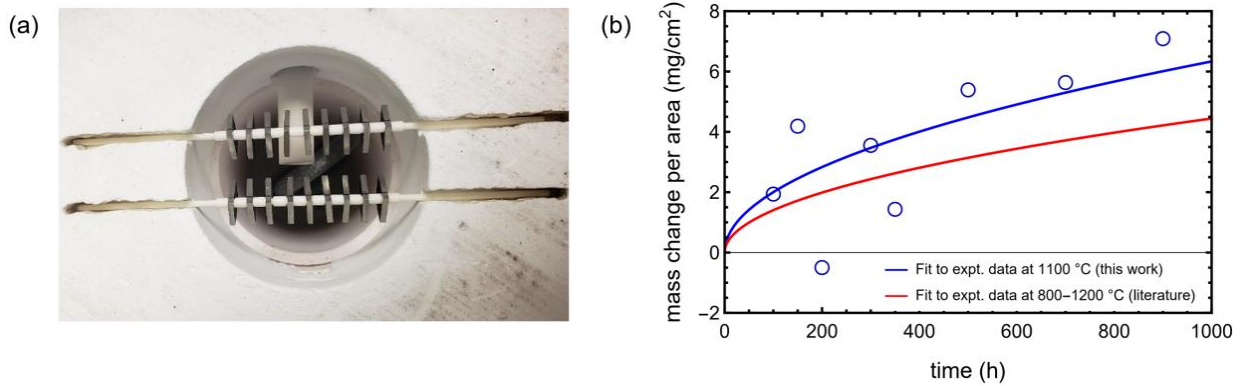

Figure 4: Corrosion experiments in $CO_2$ at 1100 °C, 2.5 MPa. (a) Image of the experimental setup showing H214 coupons sized 8 mm x 8 mm x 1.5 mm. (b) Mass change data from experiments. Each scatter point is an average of 3 measurements. The blue line is the parabolic fit to experimental data generated in this work. The red line is calculated using Equation 10a.

## Carbon Activity at the Metal-oxide | Metal Interface

Figure 5 shows interfacial carbon activity, $a_C^{int}$, computed as a function of $CO_2$ pressure at the surface, $P_{CO2}^{surf}$, and temperature, using Equations 13 and 14. Parabolic oxidation rates are calculated assuming only alumina formation on the surface. This interfacial carbon activity increases with $CO_2$ pressure and temperature. At a moderate, subcritical $CO_2$ pressure of 25 bar and high temperature of 1200 °C, the interfacial carbon activity, $a_C^{int} \sim 3.5 \times 10^{-5}$, is not sufficient to drive carbide precipitation even in the bulk of the alloy (Figure 3). However, at the highest pressure and temperature modeled (250 bar and 1200 °C), $a_C^{int}$ substantially increases to 45. Conventionally, the chemical activity of solids in their reference states (graphitic carbon for instance) is defined to be 1. In this work, activity values greater than 1 imply very strong driving forces for carburization. Metal oxidation occurs at a faster rate at higher temperatures due to the Arrhenius dependence of $k_{oxide}$ on temperature, and therefore $CO_2$ consumption and CO generation rates increase as per Equation 4. The interfacial carbon activity increases as CO partial pressure increases and $CO_2$ partial pressure decreases at the interface (Equation 11). Put together, this results in an increase in carbon activities at the interface, $a_C^{int}$, with increase in temperature. Contrary to prior purely thermodynamic analyses that indicated decreasing carbon activity with increasing temperature, these results capture more comparable trends with experimental observations [4].

At a fixed temperature, the rate of metal oxidation weakly increases with pressure (Equation 10a), whereas species diffusivities decrease with increasing pressure (Equation 7b). To attain flux balance between molecular diffusion of $CO_2$ and CO across the oxide, and their respective rates of consumption and production, reduced diffusivities at higher pressures are compensated by lower concentration of $CO_2$ and higher concentration of CO at the interface. This results in an increase in $a_C^{int}$ value with pressure.

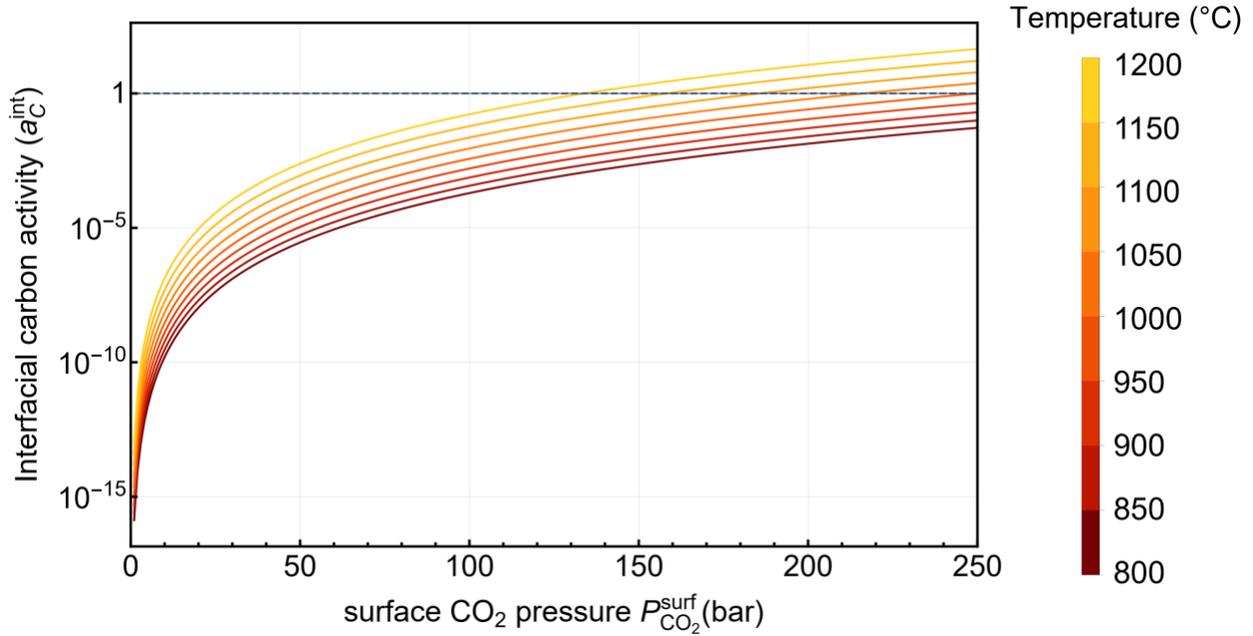

*Figure 5: Interfacial carbon activity at the metal | metal-oxide interface calculated as a function of external $CO_2$ pressure and temperature, using Equation 11. Parabolic oxidation rates in Equation 10a and baseline values for $\frac{\xi}{\epsilon}$ = 47.5 were used to calculate carbon activity. Each contour line is an isotherm with $P_{CO2}^{surf}$ up to 250 bar. There are 9 isotherms varying between 800 °C and 1200 °C in steps of 50 °C.*

The calculated carbon activity values will impact the extent of carbide precipitation in the bulk of the alloys. Sensitivity studies (Section S8 in SI) show that this carbon activity is highly sensitive to the morphological descriptor, $\xi/\epsilon$, and the parabolic oxidation rate, $k_{\text{oxide}}$ values modeled. At any pressure and temperature, carbon activity increases by up to two orders-of-magnitude when $\xi/\epsilon$, increases by a factor of 10. Larger $\xi/\epsilon$ ratios result in reduced CO and $CO_2$ diffusivity through the oxide. For fixed parabolic oxidation rates, this decrease in species diffusivity leads to CO and $CO_2$ pressures being respectively increased and decreased at the interface, which leads to higher carbon activities (Equation 11). The data shown in Figure 5 for alumina forming alloys suggests reduced possibility of carburization at pressures below 50 bar and temperatures below 1000 °C, since $a_c$ < 10-4. However, this outcome is sensitive to the magnitude of the $k_{\text{oxide}}$ and $\xi/\epsilon$ modeled, which in turn depends on the oxide and base metal composition amongst other factors. For example, $a_C^{int}$ = 5.4×10-5 at 50 bar and 1000 °C with the baseline value of $\xi/\epsilon$ = 47.5.

Increasing $\xi/\epsilon$ by roughly two order of magnitudes to 5000 changes the $a_C^{\text{int}}$ value to ~0.6 at 50 bar and 1000 °C, which is sufficient for carbide precipitation as seen in the phase diagrams in Figures 3c and 3d.

## Extent of carburization and carbide precipitate composition

Figure 6a shows the total carbide precipitate mass fractions at the metal-oxide | metal interface for MA956 and H214 at 800 °C–1100 °C, $P_{CO2}^{\text{surf}}$ = 200 bar. The interfacial carbon activities computed in Figure 5 were used as boundary conditions for these simulations. Carbon activity values that are much larger than 1, which arise for select combinations of temperature and pressure Figure 5, will result in complete carburization of the alloy, and cause numerical instabilities, and therefore not included in Figure 6. At a fixed temperature, the carbide amount at the metal | metal-oxide interface is dictated by the interfacial carbon activity, $a_C^{\text{int}}$, which is independent of time because of the assumption of quasi-steady growth rate of the oxide. The amount of carbide precipitates increases with temperature for both the MA 956 and H214 alloys, driven by the increase in $a_C^{\text{int}}$ values with temperature. However, the rate of increase in carbide mass fraction with temperature is significantly larger for the Fe-based MA956 compared to Ni-based H214. The total mass fraction of carbide phases at the Al$_2$O$_3$ | MA956 interface rises dramatically from ~0.2 at 800 °C (when $a_C^{\text{int}} = 0.013$) to ~0.78 at 1100 °C (when $a_C^{\text{int}} = 1.57$) due to the poor solubility of C in the Fe-based MA956 matrix (BCC phase). On the other hand, the carbide mass fraction in Ni-based H214 gradually increases in the 800–1000 °C range, followed by a sharper rise at 1100 °C. At 1100 °C, the carbide fraction in MA956 is nearly 4 times the value in H214.

Figure 6b shows that, consistent with trends in $a_C^{\text{int}}$, carbide precipitation decreases in both alloys with decrease in sCO$_2$ pressure, at a fixed temperature of 1100 °C. The carbide amount in MA956 is over 60% for CO$_2$ pressures over 150 bar. For the same pressure, the carbide content in H214 is below 30%. At 50 bar pressure, $a_C^{\text{int}} < 10^{-3}$, which does not create significant thermodynamic driving force for carbide formation (as also evident from Figure 3c and 3d).

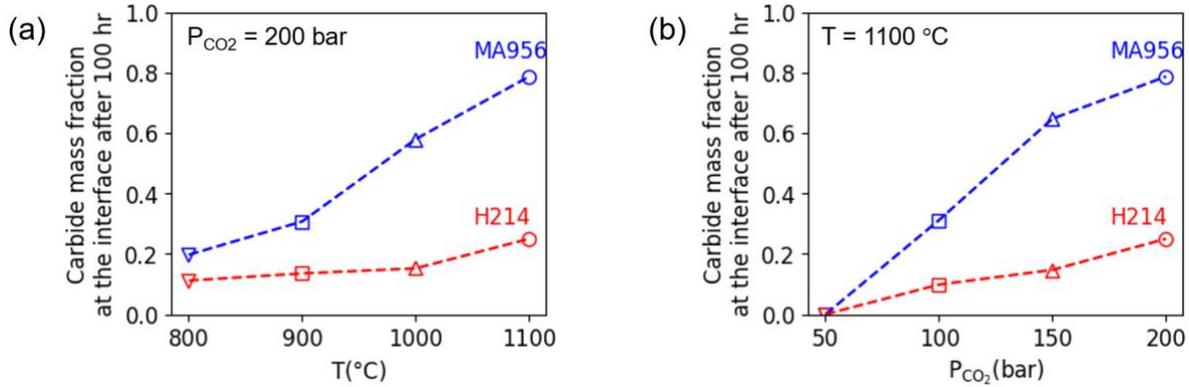

*Figure 6: (a) Total mass fraction of all carbide precipitates at the metal-oxide | metal interface at 200 bar sCO$_2$ pressure with varying temperatures for MA956 and H214. (b) (a) Total mass fraction of all carbide precipitates at the metal-oxide | metal interface at 1100 °C with varying scO$_2$ pressures for MA956 and H214. Mass fractions are computed from phase-equilibrium calculations using the DICTRA module in Thermo-Calc, with boundary conditions for carbon activity, $a_C^{\text{int}}$, obtained from the data in Figure 5. The dashed lines are visual aids.*

Figure 7 shows the spatio-temporal evolution of chemical composition of carbide phases for MA 956 (Figures 7a, c) and H214 (Figures 7b, d) when exposed to $P_{CO2}^{\text{surf}} = 200$ bar at 800 C and 1000 C. While simulation times up to 100 h is considered in this study, generally, longer time scales result in larger carbon content in the precipitates, progressing from $M_{23}C_6$ to $M_7C_3$ and $M_3C_2$, where the major metallic element M is Cr both alloys. This is because of the increase in carbon activity with temperature and faster carbon diffusion into the metal matrix. For the MA 956 at 800 °C (Figure 7a), the total carbide content at the MA956 | Al$_2$O$_3$ interface is ~20% at any time instant and the carbide phase precipitating at and near this interface is $Cr_{6.21}Fe_{0.76}X_{0.03}C_3$, where X represents other alloying elements in small concentrations. The same precipitate forms the dominant phase at depths of 0.1 mm, 0.2 mm, and 0.6 mm after simulation times of 1 h, 10 h and 100 h respectively. With increasing depth into the alloy, $Cr_{18.6}Fe_{4.2}X_{0.2}C_6$ becomes the more stable carbide phase with a mass fraction up to 15% at a depth of 0.6 mm at 100 h. This carbide phase has long tail throughout the depth of the alloy even if only present at a ~1.7% mass fraction. Increasing precipitate amounts with time result in continual depletion of the BCC phase, by up to 20% at 800 C, . At the same temperature of 800 °C, smaller extents of carbide precipitation occurs in H214 compared to MA956 (Figure 7b), with a maximum of ~10% of carbide phases by mass, and a precipitation depth at 100 h. Additionally, $Cr_{6.8}X_{0.2}C_3$ is the major precipitate phase in H214 and the precipitation depth is only ~0.25 mm after 100 h. At 1000 °C, similar trends are obtained, but with significantly larger mass fractions of the carbide precipitates for both alloys (Figures 7c and 7d). For MA 956, a mass fraction of ~60% is obtained for the $Cr_{6.21}Fe_{0.76}X_{0.03}C_3$ precipitate, and the faster diffusion of C-species at 1000 °C results in through-thickness carburization after 100 h. A distinction in the H214 alloy is the precipitation of a higher carbon content $Cr_{2.95}X_{0.05}C_2$ carbide phase, instead of $Cr_{6.8}X_{0.2}C_3$ that was observed at 800 C (Figure 7b) at a mass fraction of ~16% at the H214 | $Al_2O_3$ interface (Figure 7d). After 100 h, $Cr_{2.95}M_{0.05}C_2$ forms up to ~2 mm, and the lower carbon content, $Cr_{6.8}X_{0.2}C_3$ precipitating at greater depths within the bulk alloy. The amount of the FCC phase in H214 is ~84% at 1000 °C, compared to ~40% BCC phase in MA956 at the same temperature.

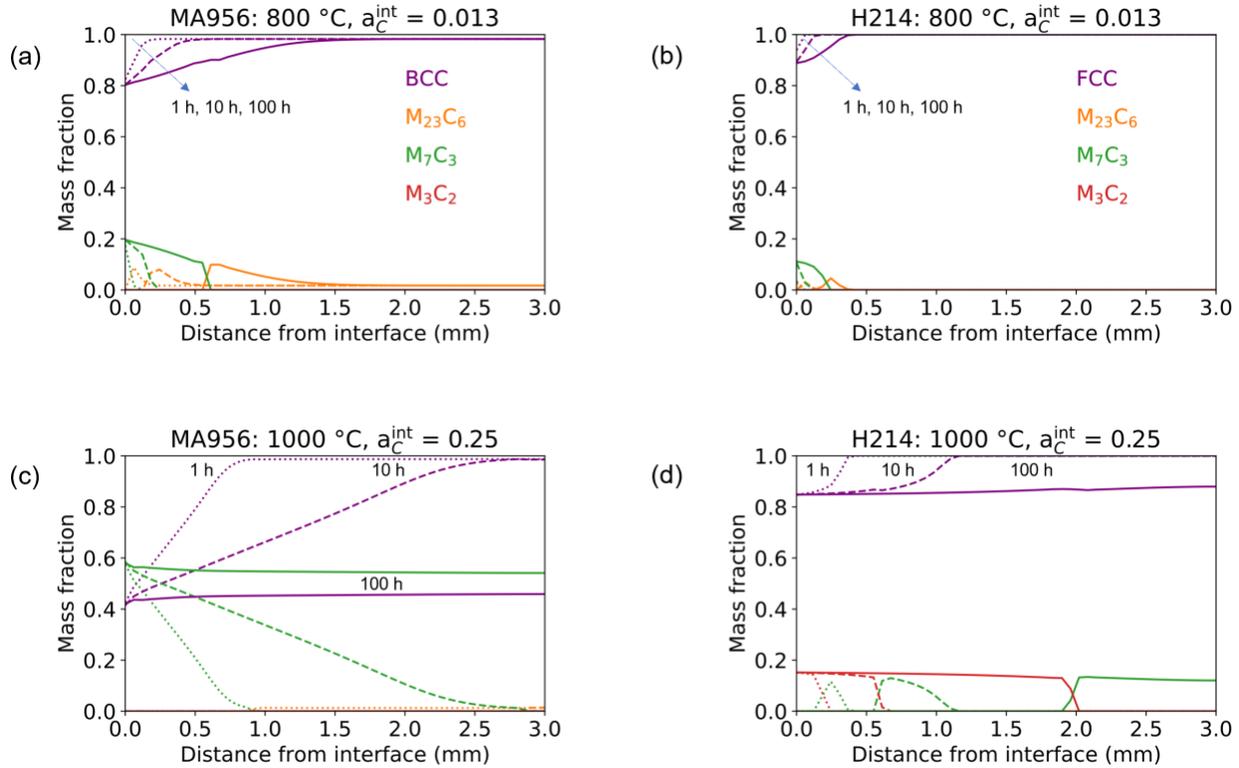

Figure 7: Carburization in MA956 (left column) and H214 (right column) from 1D diffusion calculations. (a) MA956 at 800 °C. Mass fractions of the primary alloy phase and carbide precipitates are plotted at different timestamps: 1 h (dotted lines), 10 h (dashed lines) and 100 h (solid lines). In all cases, $a_C^{int}$ is obtained from Figure 5, at $P_{CO2}^{surf} = 200$ bar (calculated using Equation 13). (b) H214 at 800 °C, (c) MA956 at 1000 °C and (d) H214 at 1000 °C. Line-styles in (b-d) follow the same pattern as in (a). Equation 10a is used to compute the parabolic oxidation rates, the baseline values for $\frac{\xi}{\epsilon} = 47.5$.

These results together with Figure 6 underscore that the Fe-based MA 956 is more susceptible to carbide precipitation compared to the Ni-based H214 at any temperature and pressure condition, largely driven by the weaker carbon solubility in the Fe compared to Ni matrix.

## 3.3 Oxide delamination due to carbide precipitation

Figure 8 shows contours of the maximum principal stress ($\sigma_1$) contours computed from the FEA model (Equation 15e) for the H214 and MA956 alloys assuming an $Al_2O_3$ oxide layer and $Cr_7C_3$ as the carbide precipitate. $Cr_7C_3$ is considered based on phase-equilibrium predictions in Figure 7. Due to the poor solubility of C in MA956, Fe is also precipitated in the carbide phases, however that is not accounted for in our calculations to simulate a worst-case scenario for carbide induced stresses. Since the elastic moduli of $Fe_7C_3$ is lower than that of $Cr_7C_3$, stresses induced due to $Fe_7C_3$ precipitation are expected to be smaller in magnitude and reduce the fracture tendency in $Al_2O_3$.

Figure 8a shows the contours of maximum principal stresses (computed in Eq. 15e) for MA956 with $Al_2O_3$ on the surface and with $Cr_7C_3$ precipitate at 800 °C, at an aspect ratio of $\frac{\phi}{\delta} = 2.5$. Stresses within the oxide layer are tensile, whereas stresses in the carbide are compressive in nature. Stresses in the metallic phase are smaller in magnitude, due to the comparatively smaller elastic modulus (Table 2). The magnitude of tensile stresses in the oxide film becomes smaller away from the triple junction interface – MA956 | $Al_2O_3$ | $Cr_7C_3$ – and towards the surface. Similarly, the magnitude of the compressive stresses in the carbide precipitate also reduces away from this interface into the bulk of the precipitate, especially near the edge with the metal matrix phase as metal can plastically deform. Stress contours are shown for H214 with $Al_2O_3$ on the surface and with $Cr_7C_3$ precipitate (Figures 8(b) – (f))) to illustrate the effects of temperature and aspect ratio of the precipitate. Compared to the MA956 results in Figure 8(a), at the same aspect ratio, the stress contours for H214 look qualitatively similar, but with relatively larger magnitudes. This is due to larger volumetric misfit strains between $Cr_7C_3$ and H214, compared to $Cr_7C_3$ and MA956 (Table 2). At 1000 °C (Figure 8(c)), the magnitude of the tensile stresses is smaller than at 800 °C (Figure 8b), because the elastic moduli of all phases decrease with an increase in temperature. Larger aspect ratios of the carbide precipitate, for a fixed lateral size, $\phi$, imply smaller volumes of the carbide phase and therefore generate smaller tensile stresses in the oxide (Figures 8d-8f). However, carbides with larger aspect ratios have sharper corners at the triple junction between the metal, oxide, and carbide phases, which generates strong compressive stresses within the carbide. For all these cases, the maximum value of $\sigma_1$, i.e., the maximum tensile stress is generated at the interface between the Cr$_7$C$_3$ carbide precipitate and Al$_2$O$_3$ oxide layer.

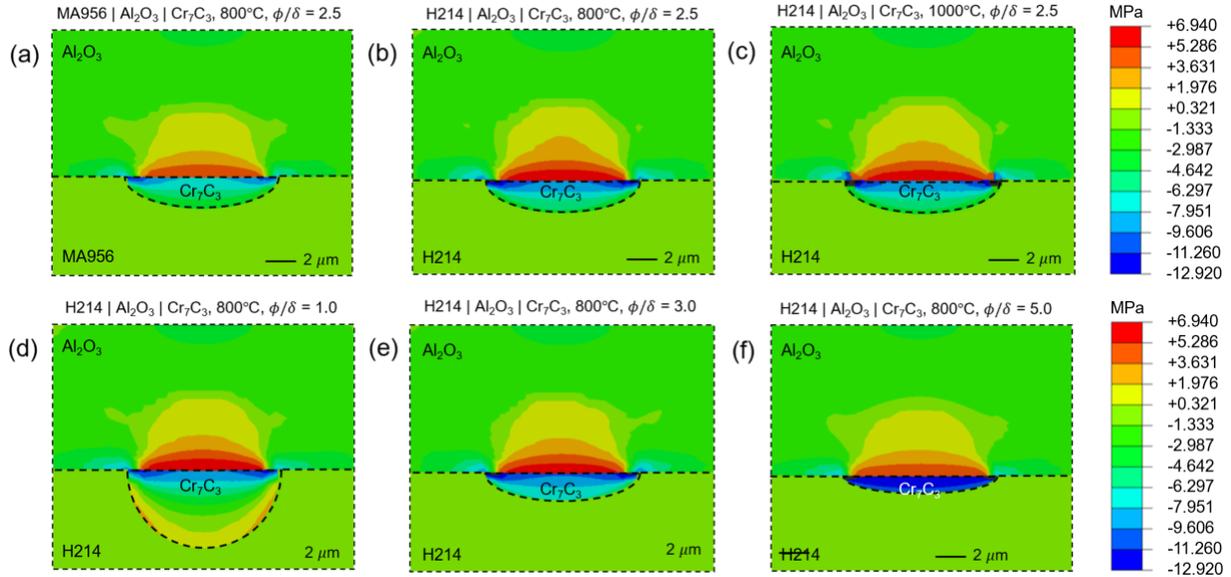

*Figure 8: Contours of the principal stress ($\sigma_1$) in (a) H214 | Al$_2$O$_3$ | Cr$_7$C$_3$ at 800 °C, (b) for H214 | Al$_2$O$_3$ | Cr$_7$C$_3$ at 800 °C and (c) H214 | Al$_2$O$_3$ | Cr$_7$C$_3$ at 1000 °C. In (a-c), the aspect ratio of the carbide was fixed at $\frac{\phi}{\delta}$ = 2.5. The black dashed lines show the geometry of the metal, metal-oxide and carbide phases. Only a magnified region near the carbide precipitate is shown here for clarity. The entire domain is depicted in Figure 1b. Values for the imposed misfit strains used to compute these stresses are reported in Table 2. Contours of $\sigma_{principal}$ in H214 | Al$_2$O$_3$ | Cr$_7$C$_3$ at 800 °C for (d) $\frac{\phi}{\delta} = 1$, (e) $\frac{\phi}{\delta} = 3$ and (f) $\frac{\phi}{\delta} = 5$. In (a-f), the lateral dimension of the carbide $\phi$ is fixed at 5 μm.*

To assess conditions that can lead to Al$_2$O$_3$ delamination, Figure 9 compares the maximum value of tensile stresses obtained from FEA analyses (Figure 8) with experimentally reported fracture stresses for Al$_2$O$_3$ films, as a function of the precipitate aspect ratio (Figure 9a) and the precipitate size (Figure 9b) for both MA 956 and H214 alloys at 800 ℃ and 1000 ℃. Carbide phases in the H214 and MA956 at each temperature were obtained from phase equilibrium predictions in Figure 7. According to the maximum stress failure criterion, brittle materials like Al$_2$O$_3$ will fail when the maximum principal stress $\sigma_1$ exceeds the tensile strength $\sigma_t$ [64]. Tensile fracture is the predominant mode of mechanical failure with tensile strength values being at least 10 times smaller than the compressive strength for alumina. Since fracture stresses can vary with type of Al$_2$O$_3$ polymorph, a range of values are shown in Figure 9a and Figure 9b, with the upper and lower bounds representing reported tensile strengths for single crystal corundum phase $\alpha$-$Al_2O_3$ (8.1 GPa) [65] and amorphous $Al_2O_3$ (4.2 GPa) [66].

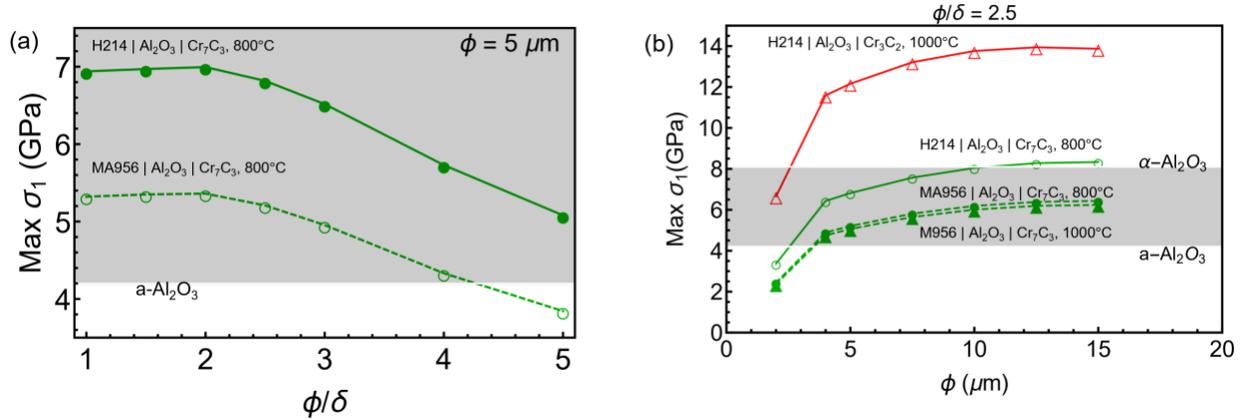

*Figure 9: (a) The maximum value of tensile principal stress $\sigma_1$ (denoted as Max $\sigma_1$) in $Al_2O_3$ for MA956 | $Al_2O_3$ | $Cr_7C_3$ and H214 | $Al_2O_3$ | $Cr_7C_3$, calculated along the oxide-carbide contact region (or interface). The lower bound of the gray band is at 4.2 GPa, which is the experimental fracture stress for amorphous $Al_2O_3$ at room temperature. (b) The maximum value of $\sigma_1$ for several metal | metal-oxide | carbide configurations identified from Figure 8. The gray horizontal band denotes experimental tensile strengths for $Al_2O_3$ at room temperature, bound by the crystalline $\alpha$-polymorph (8.1 GPa) and amorphous polymorph (4.2 GPa). The same fracture stress values are used at all considered temperatures.*

Consistent with results in Figure 8, the maximum tensile stresses are larger for H214 compared to MA956 (Figure 9a) for the same precipitate dimensions, because of the larger strain misfits between the carbides and H214 as evident in Table 2. For both alloys, the maximum stress decreases with increase in aspect ratio due to the smaller amounts of carbide precipitation in the bulk alloy. For a precipitate size of 5 $\mu$m, for all aspect ratios from 1 to 4, the maximum stresses are above the fracture limits established for $Al_2O_3$, which indicate likely fracture for both H214 and MA 956 alloys with $Cr_7C_3$ precipitates at 800 °C; for MA 956, the stress drops below the fracture limit for $\frac{\phi}{\delta} = 5$. For the same metal | metal-oxide | carbide configuration, the stress magnitudes at higher temperatures are lower due to lower elastic moduli of the ceramic phases. These trends are further illustrated in Figure 9b.

Figure 9b shows the variation in the maximum value of $\sigma_1$ (denoted as max $\sigma_1$) with precipitate size, at a fixed aspect ratio of $\phi/\delta$ = 2.5, where the stresses begin to level off in Figure 9(a). At 800°C, $Al_2O_3$ on MA956 is likely to fracture when $\phi \sim$ 3.5 $\mu$m, whereas $Al_2O_3$ on H214 is likely to fracture for even smaller precipitate sizes of $\phi \sim$ 2.5 $\mu$m. At 1000 °C, $Cr_7C_3$ is still the predominant precipitate in MA956 (shown previously in Figure 7). Since the elastic moduli of $Cr_7C_3$ decreases with increasing temperature, the value of max $\sigma_1$ for MA956 | $Al_2O_3$ | $Cr_7C_3$ is slightly lowered at 1000 °C compared to 800 °C. In contrast, the predominant precipitate in H214 at 1000 °C is $Cr_3C_2$, which is the carbide having the largest strain misfit due to the highest carbon content (detailed in Table 2). There is an even larger strain misfit between carbides $Cr_3C_2$ and H214, compared to $Cr_7C_3$ and MA 956. This results in much larger tensile stresses in the H214 | $Al_2O_3$ | $Cr_3C_2$ system. For even the lowest precipitate size modeled, $\phi \sim 2$ $\mu$m, and for an aspect ratio of 2.5, the value of max $\sigma_1$ in the H214 | $Al_2O_3$ | $Cr_3C_2$ system is already large enough to result in localized fracture of oxide layer. Although the amount of $Cr_3C_2$ in H214 is predicted to be smaller than the amount of $Cr_7C_3$ in MA956, 16 wt% in comparison with 60 wt% (Figure 7), larger volumetric misfits and therefore larger tensile stresses between carbides and H214 point to likely oxide spallation at high temperatures even for H214.

Results in Figures 8 and 9 show larger principal stress in carbides that precipitate in H214, compared to MA956 under identical precipitate morphologies. However, we know from carburization phase equilibrium modeling that H214 will have smaller extents of carbide precipitation compared to MA956 (Figures 6 and 7). Based on assumed densities of 7.06 g/cm$^3$, 6.68 g/cm$^3$, 7.25 g/cm$^3$ and 8.05 g/cm$^3$ for $Cr_7C_3$, $Cr_3C_2$, MA956 and H214, and by projecting the area of 2-D domain into the plane by a finite thickness for the metal and the precipitate, we obtain approximate estimates of equivalent precipitate dimensions for a given mass fraction. To estimate the lateral precipitate size near the metal | metal-oxide interface, an infinitesimally thin carbide precipitate is considered (such that the carbide mass fraction can be transferred to the carbide length fraction using their densities). For example, a lateral precipitate dimension of $\phi = 15$ $\mu$m translates to a mass fraction of 2.50% $Cr_3C_2$ in H214 and 2.92% $Cr_7C_3$ in MA956. However, results in Figure 7 show much higher carbide amounts at 1000 °C, which imply larger lateral precipitate dimensions. For example, the predicted mass fraction of 16 wt% for $Cr_3C_2$ in H214 at 1000 °C corresponds to a lateral precipitate dimension of $\phi = 93$ $\mu$m for a total domain width of 1000 $\mu$m. For example, the predicted mass fraction of 60 wt% for $Cr_7C_3$ in MA956 at 1000 °C corresponds to $\phi = 303$ $\mu$m for a total domain width of 1000 $\mu$m, which is more than 3 times larger than the precipitate size in H214. For these estimated precipitate sizes, the maximum tensile stress well exceeds fracture limits for both H214 and MA 956. These results and interpretations further reinforce that even with smaller extents and precipitate sizes in H214 compared to MA 956, H214 is still likely to be subjected to oxide cracking at temperatures of and exceeding 1000 °C.

The 2-D modeling domain considered for FEA analysis is only a simplified system and discounts the exact size distribution of carbide precipitates, which depends on factors such as interface energies between the matrix and precipitate, clustering effects, and alloy microstructure. These results are still useful to make comparisons between the MA956 and H214 alloy systems. Additionally, experimental reports also indicate larger precipitate sizes in Fe alloys compared to Ni alloys. For instance, $M_{23}C_6$ precipitates in Ni alloys are reported to be 25–40 nm at ~750 °C after few hours of heat treatment [67], whereas $M_{23}C_6$ precipitates in steels are reported to have dimensions on the order of 100 nm at a comparable temperature of 650 °C [68,69]. Larger precipitates are likely to introduce larger misfit stresses and promote oxide fracture in steels.

## 4. Model Limitations and Future Work

This work presents an integrated framework to understand corrosion behavior in $sCO_2$ environments by modeling carbide precipitation and its impact on oxide delamination, by developing a quasi-steady-state model that couples the experimental kinetics of oxide growth with the transport rate of reactive gaseous species through the surface oxide. While the proposed model can successfully provide quantitative insights into the corrosion behavior of metallic alloys in $sCO_2$, further improvements are required to 1) include the effects of alloying element depletion due to oxidation and 2) investigate the impact of carbide precipitate morphology (orientation, distribution, spacing and other geometric parameters) on oxide fracture. While it is possible to quantify the tensile stresses in the oxide as a function of the precipitate size, realistic precipitate dimensions depend on the base alloy microstructure. As demonstrated recently, appropriate contributions for the volumetric free energy of the carbide, interfacial free energies for the metal | carbide and oxide | carbide interfaces, strain energies can be used to develop a nucleation and growth model for carbide phases within the metal matrix [70]. Coupling these precipitate coarsening parameters with microstructural descriptors (such as grain size distribution, grain boundary nucleation energy, etc.) can provide additional insights into the morphology of carbide precipitates.

## 5. Conclusion

In summary, the corrosion performance of model alloys MA956 and H214 in $CO_2$ atmospheres for temperatures ranging from 800–1200 °C and pressures up to 250 bar has been computationally investigated. This framework integrates thermodynamic calculations with kinetic simulations and finite-element-method-based stress field simulations to develop physical insights into $CO_2$-induced corrosion in metallic alloys. The first key contribution of this work is the development of a model to bridge data from corrosion experiments with the thermodynamics of the $CO_2$ dissociation equilibria, which is used to compute driving forces for carburization at the metal | metal-oxide interface under exposure to $CO_2$ environments. Parabolic oxidation rates for the formation of alumina on metal surfaces was deduced from reported experimental data for time-dependent mass-change of Fe- and Ni- alloys when exposed to oxidizing conditions. Data-driven fits for oxidation constants as a function of temperature and pressured matched within an order of magnitude to in-house experimental measurements for H214 coupons at 1100 °C. CALPHAD-based thermodynamic and kinetic calculations showed that H214 is more resistant to carburization than MA956 at temperatures below 900 °C, due to smaller extent of $Cr_7C_3$ precipitation (lower mass fraction and carburization depth). At higher temperatures, through-thickness carbide formation in both alloys (primarily $Cr_7C_3$ in MA956 and $Cr_3C_2$ in H214) suggests severe corrosion. Second, finite element method-based calculations were performed to investigate the impact of carbide precipitates on the mechanical stability of the oxide film. Oxide fracture in these conditions is dictated by two competing effects – (1) volumetric misfit magnitudes for alumina and carbide precipitates in the H214 and MA956 matrix phases and (2) the size of carbide precipitates formed. While tensile stresses induced in alumina on H214 are larger than tensile stresses induced in alumina on MA956, the amount of carbide precipitation is larger in MA956. According to the maximum stress criterion for brittle failure, our results indicate strong likelihood of $Al_2O_3$ fracture for a lateral precipitate dimension larger than 10 $\mu$m and depth ~2 $\mu$m. Larger misfit strains between $Cr_3C_2$ (which is the carbide with the largest carbon content and therefore largest volumetric misfit) and H214 generate larger tensile stresses in the H214 | $Al_2O_3$ | $Cr_3C_2$ system.

The systematic integration of various physical effects including material thermodynamics, reaction kinetics, and stress evolution presents a promising approach to understand and evaluate the performance of metallic alloys in oxidizing and carburizing atmospheres. This will enable the development of materials for next-generation energy infrastructure and to operate in extreme high-temperature and corrosive environments.

## Author Contributions

A.S. and R.B.C. conceived the research and wrote and revised most of the manuscript; A.S. led theoretical/computational model development and generated majority of the plots, and analyzed/interpreted data; A.S. and R.B.C. formulated and developed kinetic models for oxidation and carburization; A.S. and L.Q. formulated and developed oxide delamination theory/modeling; A.F. and J.P. developed the experimental corrosion rig to obtain data for the oxidation rate constant; A.F. wrote/revised the sections pertinent to these experiments; B.K. performed supplementary calculations in ThermoCalc to evaluate carbide mass fractions in Figure 6; R.B.C secured funding and supervised the project.

## Declaration of Interests

There are no competing interests to declare.

## Acknowledgements

This work was supported by the Department of Energy Advanced Research Projects Agency-Energy (ARPA-E) under co-operative agreement DE-AR0001123 with Michigan State University and the University of Michigan through the HITEMMP program. This research was supported in part through computational resources and services provided by Advanced Research Computing (ARC), a division of Information and Technology Services (ITS) at the University of Michigan, Ann Arbor. We acknowledge discussions with other team members from Michigan State University including Andre Benard, Himanshu Sahasrabudhe, Haseung Chung, and Patrick Kwon.

## Data Availability

The raw data required to reproduce these findings can be shared upon reasonable request from the lead and corresponding authors of this study.

# References

[1] Kruizenga A, Fleming D. Materials Corrosion Concerns for Supercritical Carbon Dioxide Heat Exchangers. Volume 3B: Oil and Gas Applications; Organic Rankine Cycle Power Systems; Supercritical CO2 Power Cycles; Wind Energy, American Society of Mechanical Engineers; 2014. https://doi.org/10.1115/GT2014-26061.

[2] Iverson BD, Conboy TM, Pasch JJ, Kruizenga AM. Supercritical CO2 Brayton cycles for solar-thermal energy. Appl Energy 2013;111:957–70. https://doi.org/10.1016/j.apenergy.2013.06.020.

[3] Turchi CS, Ma Z, Dyreby J. Supercritical Carbon Dioxide Power Cycle Configurations for Use in Concentrating Solar Power Systems. Volume 5: Manufacturing Materials and Metallurgy; Marine; Microturbines and Small Turbomachinery; Supercritical CO2 Power Cycles, American Society of Mechanical Engineers; 2012, p. 967–73. https://doi.org/10.1115/GT2012-68932.

[4] Pint BA, Pillai RR. Lifetime Model Development for Supercritical CO2 CSP Systems. Oak Ridge, TN (United States): 2019. https://doi.org/10.2172/1515655.

[5] Cui G, Yang Z, Liu J, Li Z. A comprehensive review of metal corrosion in a supercritical CO2 environment. International Journal of Greenhouse Gas Control 2019;90:102814. https://doi.org/10.1016/j.ijggc.2019.102814.

[6] Rouillard F, Moine G, Tabarant M, Ruiz JC. Corrosion of 9Cr Steel in CO2 at Intermediate Temperature II: Mechanism of Carburization. Oxidation of Metals 2012;77:57–70. https://doi.org/10.1007/s11085-011-9272-4.

[7] Wang Y, Liu Y, Tang H, Li W. Oxidation behaviors of porous Haynes 214 alloy at high temperatures. Mater Charact 2015;107:283–92. https://doi.org/10.1016/j.matchar.2015.07.026.

[8] García-Alonso MC, González-Carrasco JL, Escudero ML, Chao J. Oxidation Behavior of Fine-Grain MA 956 Superalloy. Oxidation of Metals 2000;53:77–98. https://doi.org/10.1023/A:1004582713929.

[9] Young DJ, Chyrkin A, He J, Grüner D, Quadakkers WJ. Slow Transition from Protective to Breakaway Oxidation of Haynes 214 Foil at High Temperature. Oxidation of Metals 2013;79:405–27. https://doi.org/10.1007/s11085-013-9364-4.

[10] Firouzdor V, Sridharan K, Cao G, Anderson M, Allen TR. Corrosion of a stainless steel and nickel-based alloys in high temperature supercritical carbon dioxide environment. Corros Sci 2013;69:281–91. https://doi.org/10.1016/j.corsci.2012.11.041.

[11] Olivares RI, Young DJ, Marvig P, Stein W. Alloys SS316 and Hastelloy-C276 in Supercritical CO2 at High Temperature. Oxidation of Metals 2015;84:585–606. https://doi.org/10.1007/s11085-015-9589-5.

[12] Chen H, Kim SH, Long C, Kim C, Jang C. Oxidation behavior of high-strength FeCrAl alloys in a high-temperature supercritical carbon dioxide environment. Progress in Natural Science: Materials International 2018;28:731–9. https://doi.org/10.1016/j.pnsc.2018.11.004.

[13] Holcomb GR, Carney C, Doğan ÖN. Oxidation of alloys for energy applications in supercritical CO2 and H2O. Corros Sci 2016;109:22–35. https://doi.org/10.1016/j.corsci.2016.03.018.

[14] Young DJ, Zhang J. Alloy Corrosion by Hot CO2 Gases. JOM 2018;70:1493–501. https://doi.org/10.1007/s11837-018-2944-7.

[15] Liang Z, Gui Y, Zhao Q. High-temperature corrosion resistance of nickel-base alloy 617 in supercritical carbon dioxide environment. Mater Res Express 2020;7:016548. https://doi.org/10.1088/2053-1591/ab6388.

[16] Yan C, Zhengdong L, Godfrey A, Wei L, Yuqing W. Microstructure evolution and mechanical properties of Inconel 740H during aging at 750 °C. Materials Science and Engineering: A 2014;589:153–64. https://doi.org/10.1016/j.msea.2013.09.076.

[17] Sridharan K. Corrosion of Structural Materials for Advanced Supercritical Carbon- Dioxide Brayton Cycle. Idaho Falls, ID (United States): 2017. https://doi.org/10.2172/1358349.

[18] Chen H, Kim SH, Jang C. Effect of high-temperature supercritical carbon dioxide exposure on the microstructure and tensile properties of diffusion-bonded Alloy 690. J Mater Sci 2020;55:3652–67. https://doi.org/10.1007/s10853-019-04222-z.

[19] Brittan AM, Mahaffey JT, Colgan NE, Elbakhshwan M, Anderson MH. Carburization resistance of cu-coated stainless steel in supercritical carbon dioxide environments. Corros Sci 2020;169:108639. https://doi.org/10.1016/j.corsci.2020.108639.

[20] Young DJ, Nguyen TD, Felfer P, Zhang J, Cairney JM. Penetration of protective chromia scales by carbon. Scr Mater 2014;77:29–32. https://doi.org/10.1016/j.scriptamat.2014.01.009.

[21] Gheno T, Monceau D, Zhang J, Young DJ. Carburisation of ferritic Fe–Cr alloys by low carbon activity gases. Corros Sci 2011;53:2767–77. https://doi.org/10.1016/j.corsci.2011.05.013.

[22] Gong Y, Young DJ, Kontis P, Chiu YL, Larsson H, Shin A, et al. On the breakaway oxidation of Fe9Cr1Mo steel in high pressure CO2. Acta Mater 2017;130:361–74. https://doi.org/10.1016/j.actamat.2017.02.034.

[23] Adam B, Teeter L, Mahaffey J, Anderson M, Árnadóttir L, Tucker JD. Effects of Corrosion in Supercritical CO2 on the Microstructural Evolution in 800H Alloy. Oxidation of Metals 2018;90:453–68. https://doi.org/10.1007/s11085-018-9852-7.

[24] Chang YA, Oates WA. Materials Thermodynamics. Hoboken, NJ, USA: John Wiley & Sons, Inc.; 2009. https://doi.org/10.1002/9780470549940.

[25] Gibbs GB. A model for mild steel oxidation in CO2. Oxidation of Metals 1973;7:173–84. https://doi.org/10.1007/BF00610578.

[26] Wolf I, Grabke HJ. A study on the solubility and distribution of carbon in oxides. Solid State Commun 1985;54:5–10. https://doi.org/10.1016/0038-1098(85)91021-X.

[27] Young DJ. Simultaneous oxidation and carburisation of chromia forming alloys. Int J Hydrogen Energy 2007;32:3763–9. https://doi.org/10.1016/j.ijhydene.2006.08.027.

[28] Robertson J, Manning MI. Criteria for formation of single layer, duplex, and breakaway scales on steels. Materials Science and Technology 1988;4:1064–71. https://doi.org/10.1179/mst.1988.4.12.1064.

[29] Fischer FD, Svoboda J, Antretter T, Kozeschnik E. Stress relaxation by power-law creep during growth of a misfitting precipitate. Int J Solids Struct 2016;96:74–80. https://doi.org/10.1016/j.ijsolstr.2016.06.021.

[30] Joy JK, Umale T, Zhao D, Solomou A, Xie K, Karaman I, et al. Effects of microstructure and composition on constitutive response of high temperature shape memory alloys: micromechanical modeling using 3-D reconstructions with experimental validation. Acta Mater 2022:117929. https://doi.org/10.1016/j.actamat.2022.117929.

[31] Nibennaoune Z, George D, Ahzi S, Ruch D, Remond Y, Gracio JJ. Numerical simulation of residual stresses in diamond coating on Ti-6Al-4V substrate. Thin Solid Films 2010;518:3260–6. https://doi.org/10.1016/j.tsf.2009.12.092.

[32] Akhlaghi M, Steiner T, Meka SR, Mittemeijer EJ. Misfit-induced changes of lattice parameters in two-phase systems: coherent/incoherent precipitates in a matrix. J Appl Crystallogr 2016;49:69–77. https://doi.org/10.1107/S1600576715022608.

[33] Moridi A, Ruan H, Zhang LC, Liu M. Residual stresses in thin film systems: Effects of lattice mismatch, thermal mismatch and interface dislocations. Int J Solids Struct 2013;50:3562–9. https://doi.org/10.1016/j.ijsolstr.2013.06.022.

[34] Hawa HA el, Bhattacharyya A, Maurice D. Modeling of thermal and lattice misfit stresses within a thermal barrier coating. Mechanics of Materials 2018;122:159–70. https://doi.org/10.1016/j.mechmat.2018.03.009.

[35] Al-Athel K, Loeffel K, Liu H, Anand L. Modeling decohesion of a top-coat from a thermally-growing oxide in a thermal barrier coating. Surf Coat Technol 2013;222:68–78. https://doi.org/10.1016/j.surfcoat.2013.02.005.

[36] Bergsmo A, Dunne FPE. Competing mechanisms of particle fracture, decohesion and slip-driven fatigue crack nucleation in a PM nickel superalloy. Int J Fatigue 2020;135:105573. https://doi.org/10.1016/j.ijfatigue.2020.105573.

[37] Forschelen PJJ, Suiker ASJ, van der Sluis O. Effect of residual stress on the delamination response of film-substrate systems under bending. Int J Solids Struct 2016;97–98:284–99. https://doi.org/10.1016/j.ijsolstr.2016.07.020.

[38] Inconel 956 property sheet. Http://SpecialmetalsIr/Images/Technical_info/Fer-Base/Incoloy-Alloy-MA956Pdf n.d.

[39] Haynes 214 property sheet. Https://HaynesintlCom/Docs/Default-Source/Pdfs/New-Alloy-Brochures/High-Temperature-Alloys/Brochures/214-BrochurePdf?Sfvrsn=bf7229d4_30 n.d.

[40] Kattner UR. THE CALPHAD METHOD AND ITS ROLE IN MATERIAL AND PROCESS DEVELOPMENT. Tecnol Metal Mater Min 2016;13:3–15. https://doi.org/10.4322/2176-1523.1059.

[41] Andersson J-O, Helander T, Höglund L, Shi P, Sundman B. Thermo-Calc & DICTRA, computational tools for materials science. Calphad 2002;26:273–312. https://doi.org/10.1016/S0364-5916(02)00037-8.

[42] TCdatabase. Https://ThermocalcCom/Content/Uploads/Brochures_and_Flyers/Current/Marketing-Database-OverviewPdf n.d.

[43] Allam IM. Carburization/Oxidation behavior of alloy haynes-214 in methane-hydrogen gas mixtures. Oxidation of Metals 2009;72. https://doi.org/10.1007/s11085-009-9151-4.

[44] Zhang H, Chen J, Zhang J. Performance analysis and parametric study of a solid oxide fuel cell fueled by carbon monoxide. Int J Hydrogen Energy 2013;38:16354–64. https://doi.org/10.1016/j.ijhydene.2013.09.144.

[45] Nguyen TD, la Fontaine A, Yang L, Cairney JM, Zhang J, Young DJ. Atom probe study of impurity segregation at grain boundaries in chromia scales grown in CO2 gas. Corros Sci 2018;132:125–35. https://doi.org/10.1016/j.corsci.2017.12.024.

[46] Pint BA, Pillai R, Keiser JR. Effect of Supercritical CO2 on Steel Ductility at 450°-650°C. Volume 10: Supercritical CO2, American Society of Mechanical Engineers; 2021. https://doi.org/10.1115/GT2021-59383.

[47] Furukawa T, Inagaki Y, Aritomi M. Compatibility of FBR structural materials with supercritical carbon dioxide. Progress in Nuclear Energy 2011;53:1050–5. https://doi.org/10.1016/j.pnucene.2011.04.030.

[48] Reaction-Web - properties of a species or chemical reaction. Https://WwwCrctPolymtl.ca/ReacwebHtm n.d.

[49] Kontogeorgis GM, Privat R, Jaubert J-N. Taking Another Look at the van der Waals Equation of State–Almost 150 Years Later. J Chem Eng Data 2019;64:4619–37. https://doi.org/10.1021/acs.jced.9b00264.

[50] Borgenstam A, Höglund L, Ågren J, Engström A. DICTRA, a tool for simulation of diffusional transformations in alloys. Journal of Phase Equilibria 2000;21:269–80. https://doi.org/10.1361/105497100770340057.

[51] Jain A, Ong SP, Hautier G, Chen W, Richards WD, Dacek S, et al. Commentary: The Materials Project: A materials genome approach to accelerating materials innovation. APL Mater 2013;1:011002. https://doi.org/10.1063/1.4812323.

[52] Li X, Ma Y, Zhou W, Liu Z. Hydrogen Atom and Molecule Adsorptions on FeCrAl (100) Surface: A First-Principle Study. Front Energy Res 2021;9. https://doi.org/10.3389/fenrg.2021.713493.

[53] Samin AJ, Taylor CD. First-principles investigation of surface properties and adsorption of oxygen on Ni-22Cr and the role of molybdenum. Corros Sci 2018;134:103–11. https://doi.org/10.1016/j.corsci.2018.02.017.

[54] Tolpygo VK, Clarke DR. Wrinkling of α-alumina films grown by thermal oxidation—I. Quantitative studies on single crystals of Fe–Cr–Al alloy. Acta Mater 1998;46:5153–66. https://doi.org/10.1016/S1359-6454(98)00133-5.

[55] Smith M. ABAQUS Standard User's Manual, Version 6.9 2009.

[56] FeCrAl plasticity. Https://MooseframeworkInlGov/Bison/Source/Materials/Tensor_mechanics/FeCrAlPowerLawHardeningStressUpdateHtml n.d.

[57] Latella BA, Liu T. High-Temperature Young's Modulus of Alumina During Sintering. Journal of the American Ceramic Society 2005;88:773–6. https://doi.org/10.1111/j.1551-2916.2005.00082.x.

[58] FeCrAl modulus. Https://MooseframeworkInlGov/Bison/Source/Materials/Tensor_mechanics/FeCrAlElasticityTensorHtml n.d.

[59] NiCrAl modulus. Https://WwwHaynesintlCom/Alloys/Alloy-Portfolio_/High-Temperature-Alloys/Haynes-r-41-Alloy/Elastic-Modulus n.d.

[60] Sun L, Ji X, Zhao L, Zhai W, Xu L, Dong H, et al. First Principles Investigation of Binary Chromium Carbides Cr7C3, Cr3C2 and Cr23C6: Electronic Structures, Mechanical Properties and Thermodynamic Properties under Pressure. Materials 2022;15:558. https://doi.org/10.3390/ma15020558.

[61] Gong X, Cui C, Yu Q, Wang W, Xu W-W, Chen L. First-principles study of phase stability and temperature-dependent mechanical properties of (Cr, M)23C6 (M = Fe, Mo) phases. J Alloys Compd 2020;824:153948. https://doi.org/10.1016/j.jallcom.2020.153948.

[62] Li W, Wang R, Li D, Fang D. A Model of Temperature-Dependent Young's Modulus for Ultrahigh Temperature Ceramics. Physics Research International 2011;2011:1–3. https://doi.org/10.1155/2011/791545.

[63] Xu C, Gao W. Pilling-Bedworth ratio for oxidation of alloys. Materials Research Innovations 2000;3:231–5. https://doi.org/10.1007/s100190050008.

[64] Li S. The Maximum Stress Failure Criterion and the Maximum Strain Failure Criterion: Their Unification and Rationalization. Journal of Composites Science 2020;4:157. https://doi.org/10.3390/jcs4040157.

[65] Konstantiniuk F, Tkadletz M, Czettl C, Schalk N. Fracture Properties of α– and κ–Al2O3 Hard Coatings Deposited by Chemical Vapor Deposition. Coatings 2021;11:1359. https://doi.org/10.3390/coatings11111359.

[66] Frankberg EJ, Kalikka J, García Ferré F, Joly-Pottuz L, Salminen T, Hintikka J, et al. Highly ductile amorphous oxide at room temperature and high strain rate. Science (1979) 2019;366:864–9. https://doi.org/10.1126/science.aav1254.

[67] Garosshen TJ, McCarthy GP. Low temperature carbide precipitation in a nickel base superalloy. Metallurgical Transactions A 1985;16:1213–23. https://doi.org/10.1007/BF02670326.

[68] Tioguem F, N'guyen F, Mazière M, Tankoua F, Galtier A, Gourgues Lorenzon A-F. Advanced quantification of the carbide spacing and correlation with dimple size in a high-strength medium carbon martensitic steel. Mater Charact 2020;167:110531. https://doi.org/10.1016/j.matchar.2020.110531.

[69] Prat O, Garcia J, Rojas D, Carrasco C, Kaysser-Pyzalla AR. Investigations on coarsening of MX and M23C6 precipitates in 12% Cr creep resistant steels assisted by computational thermodynamics. Materials Science and Engineering: A 2010;527:5976–83. https://doi.org/10.1016/j.msea.2010.05.084.

[70] Murty TN, Singh RN, Ståhle P. Delayed hydride cracking of Zr-2.5%Nb pressure tube material due to partially constrained precipitates. Journal of Nuclear Materials 2019;513:129–42. https://doi.org/10.1016/j.jnucmat.2018.10.040.